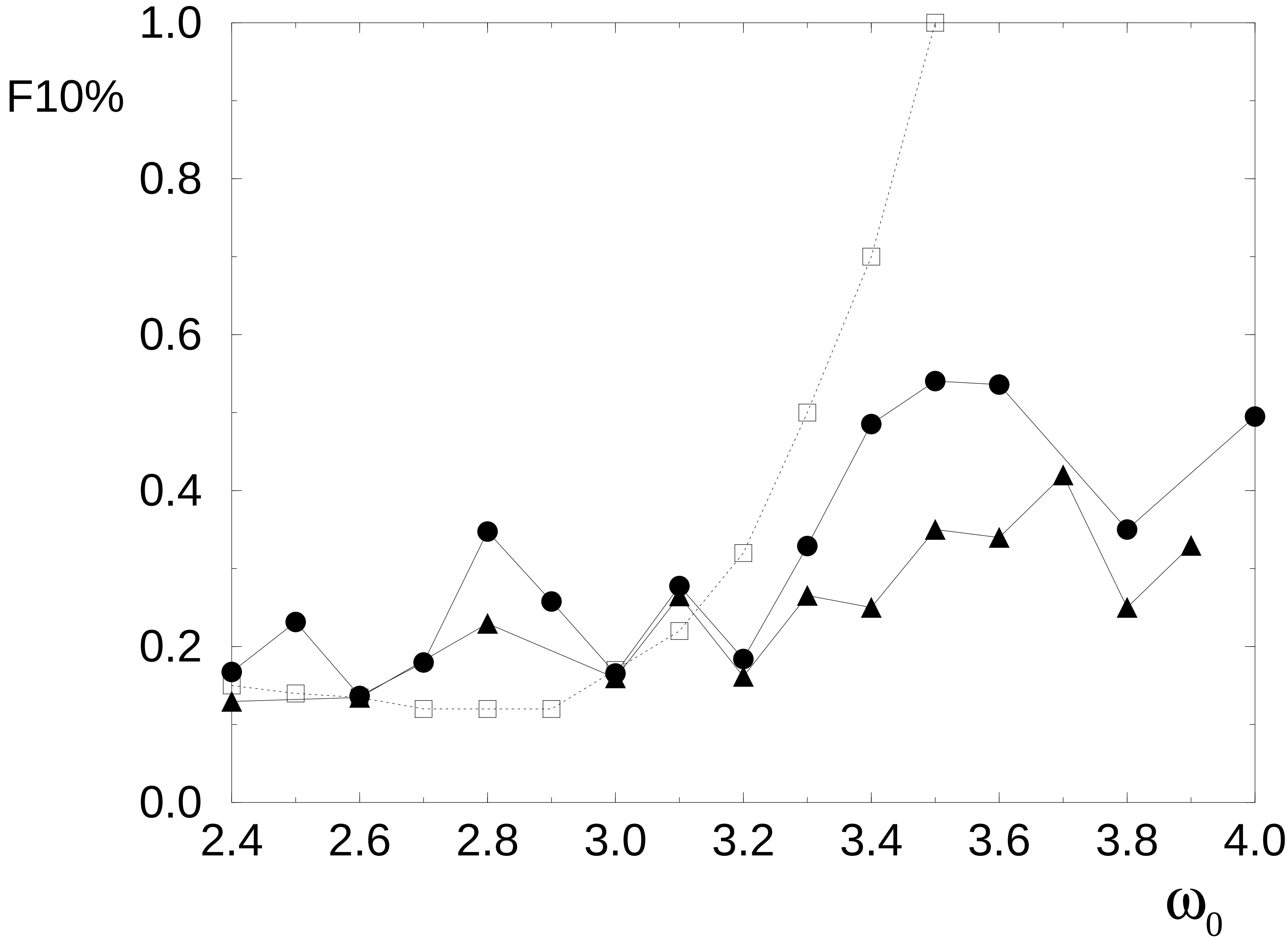
F10%
1.0
0.8
0.6
0.4
0.2
0.0
2.4
2.6
2.8
3.0
3.2
3.4
3.6
3.8
4.0
$\omega_0$

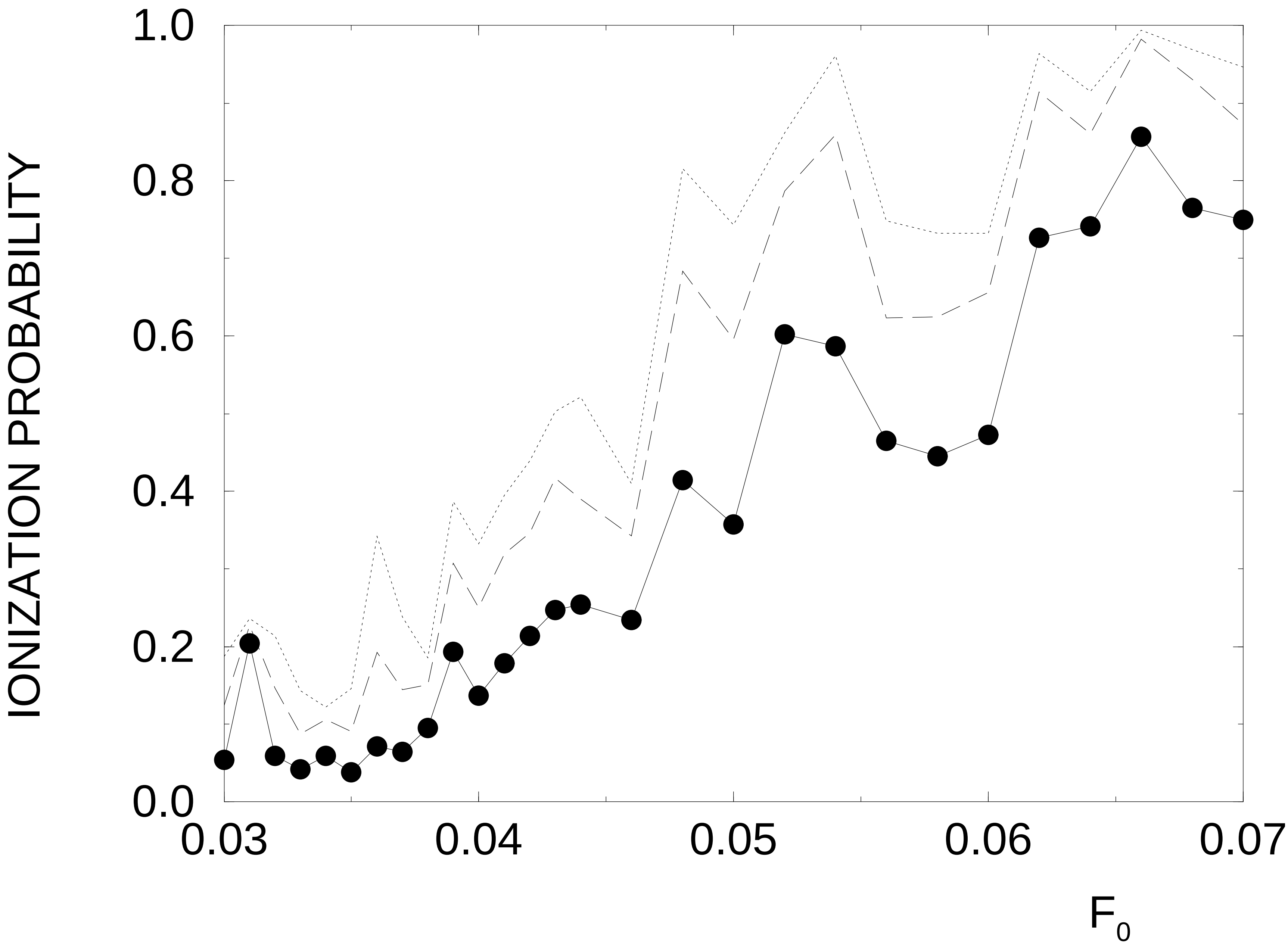

IONIZATION PROBABILITY
1.0
0.8
0.6
0.4
0.2
0.0
0.03
0.04
0.05
0.06
0.07
$F_0$

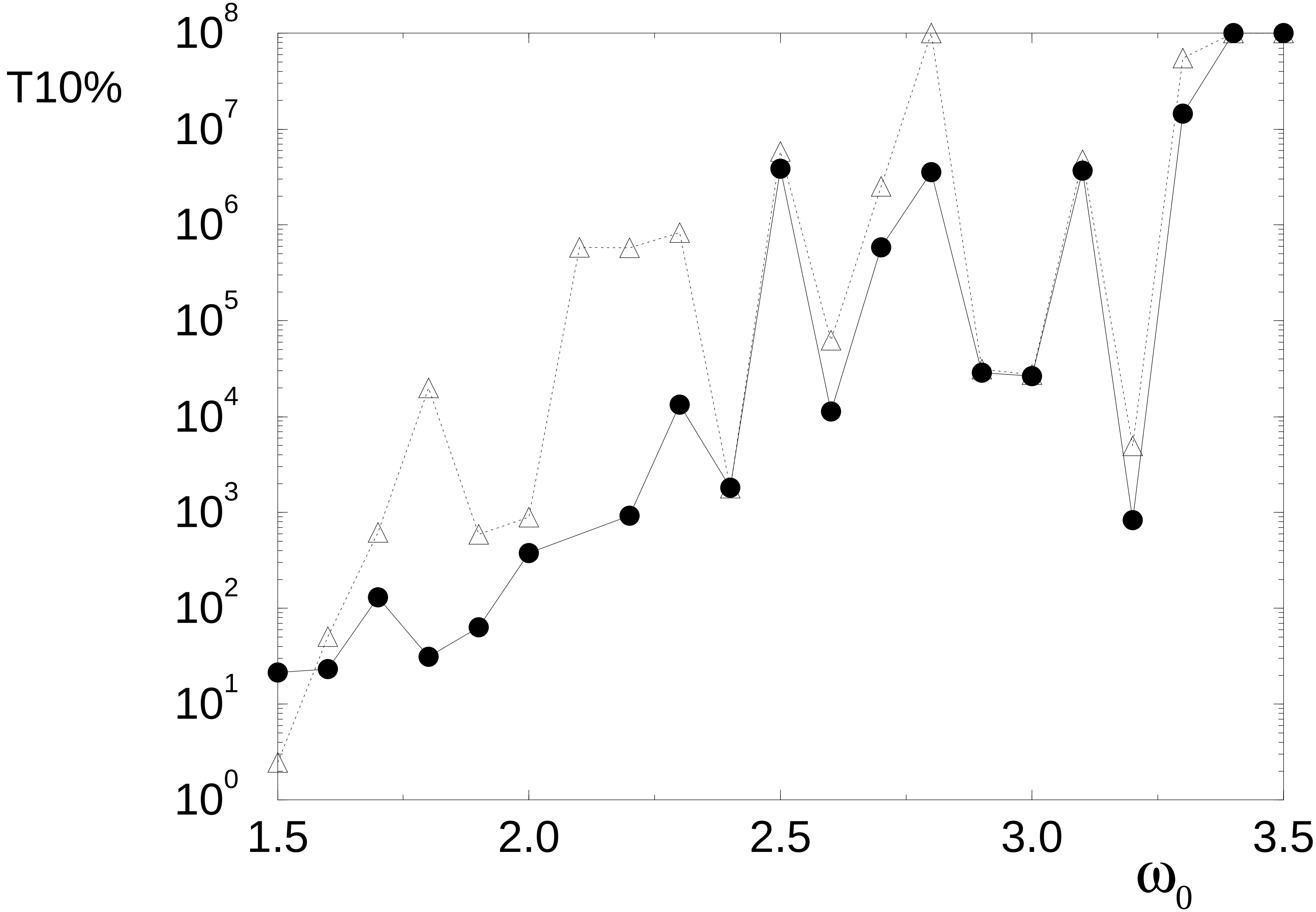

T10%
$10^8$
$10^7$
$10^6$
$10^5$
$10^4$
$10^3$
$10^2$
$10^1$
$10^0$
1.5
2.0
2.5
3.0
3.5
$\omega_0$

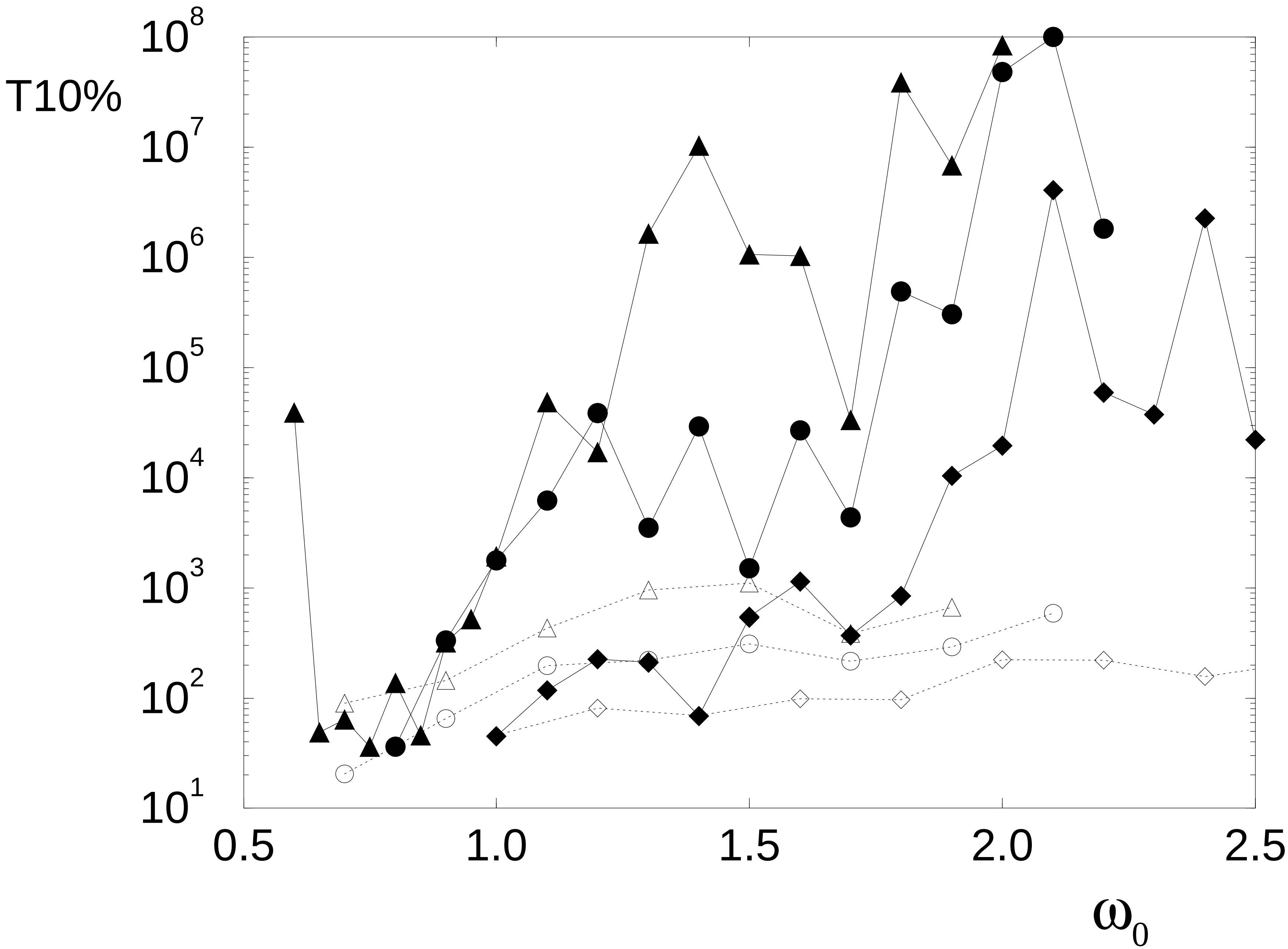

T10%
10^8
10^7
10^6
10^5
10^4
10^3
10^2
10^1
0.5
1.0
1.5
2.0
2.5
ω0

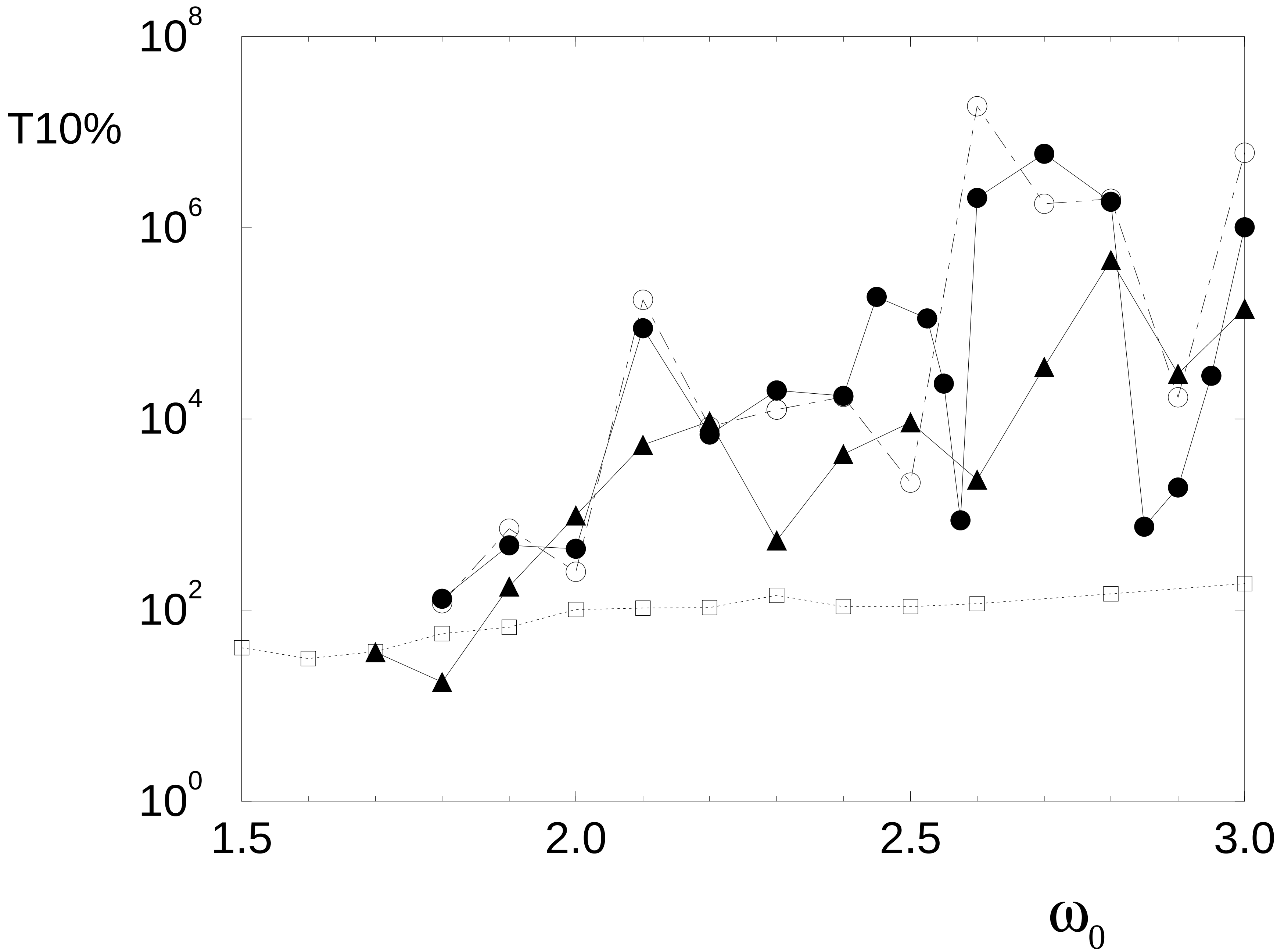

T10%
$10^8$
$10^6$
$10^4$
$10^2$
$10^0$
1.5
2.0
2.5
3.0
$\omega_0$

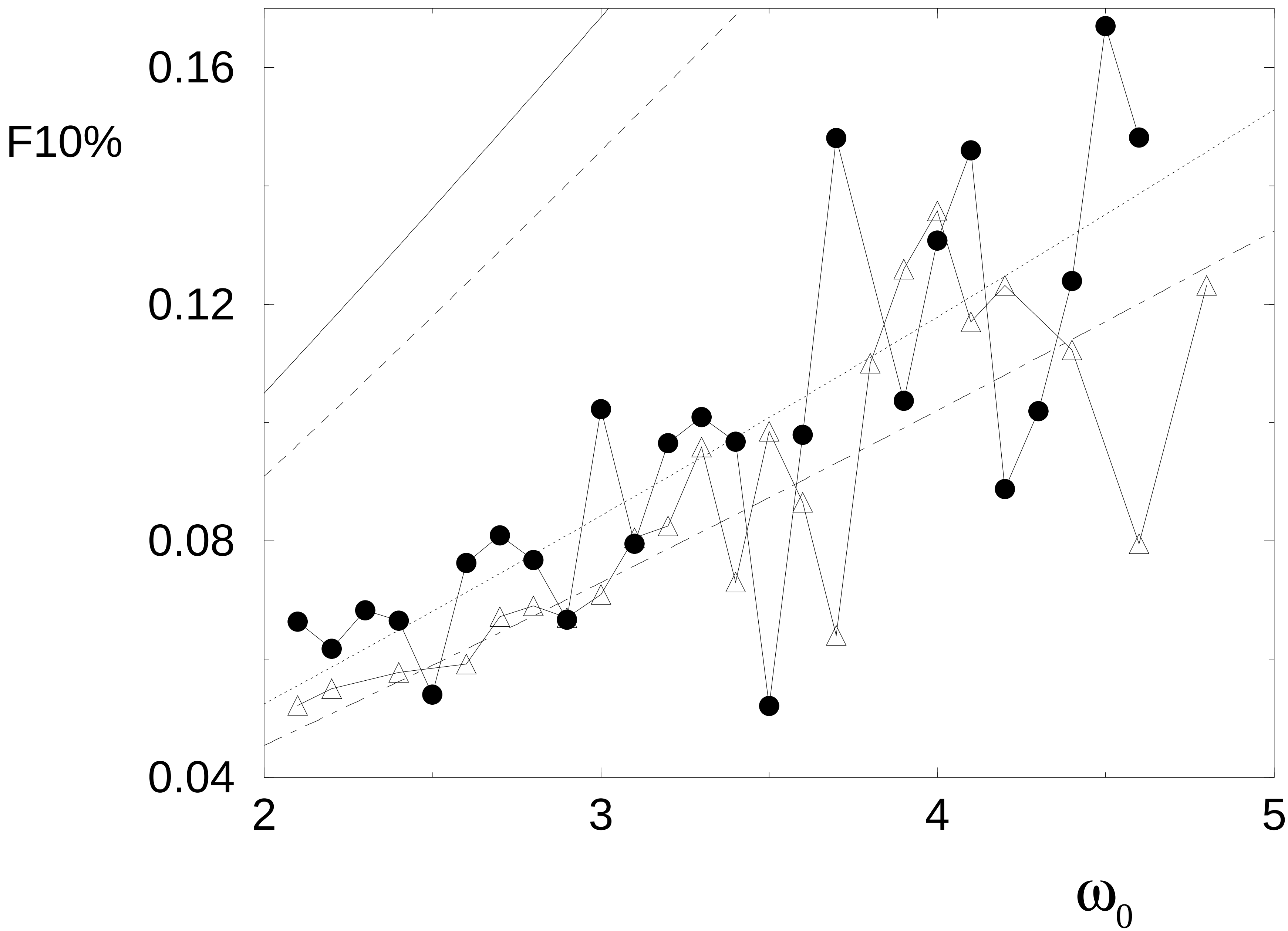

F10%
0.16
0.12
0.08
0.04
2
3
4
5
ω0

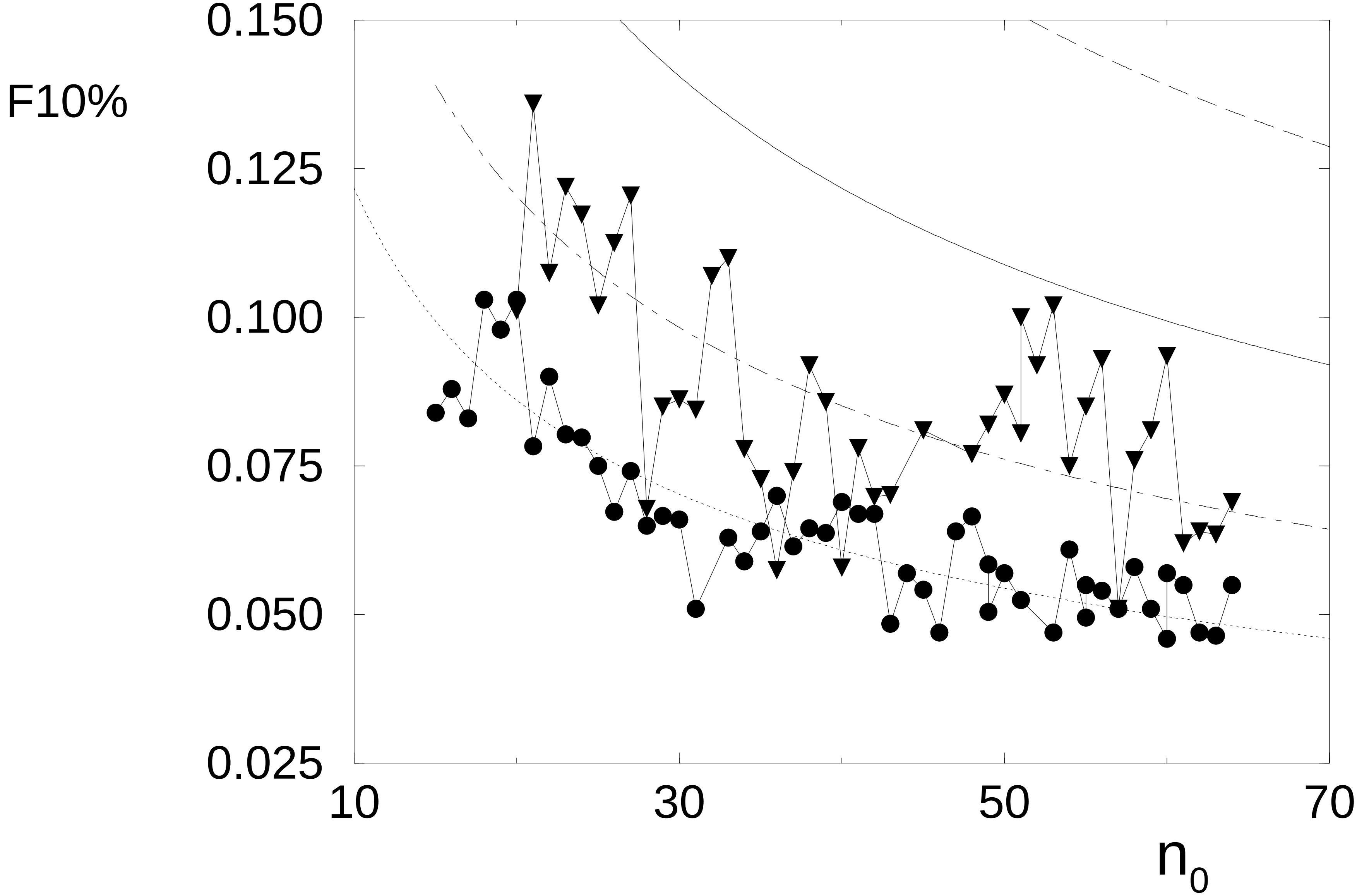

F10%
0.150
0.125
0.100
0.075
0.050
0.025
10
30
50
70
$n_0$

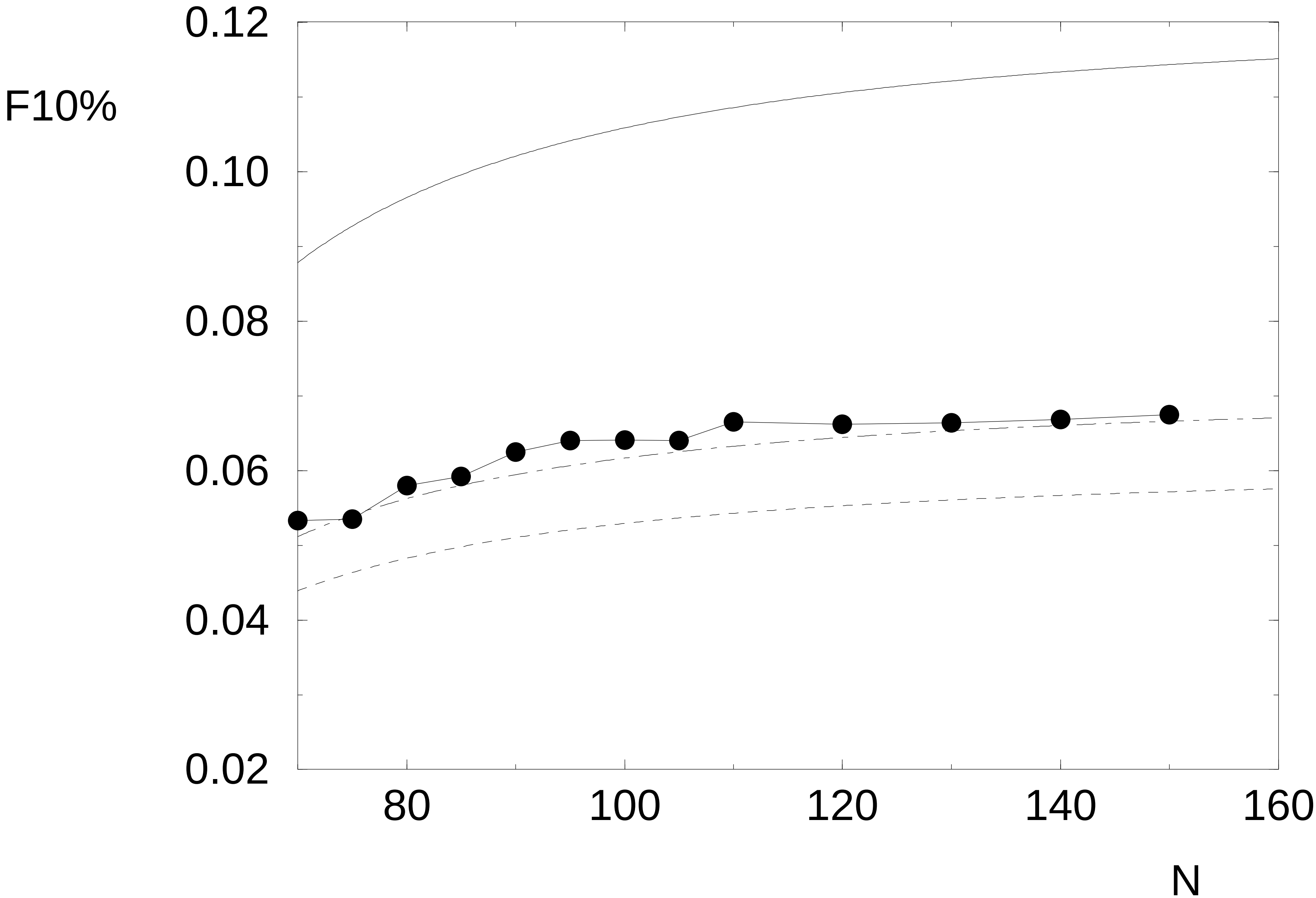

F10%
0.12
0.10
0.08
0.06
0.04
0.02
80
100
120
140
160
N

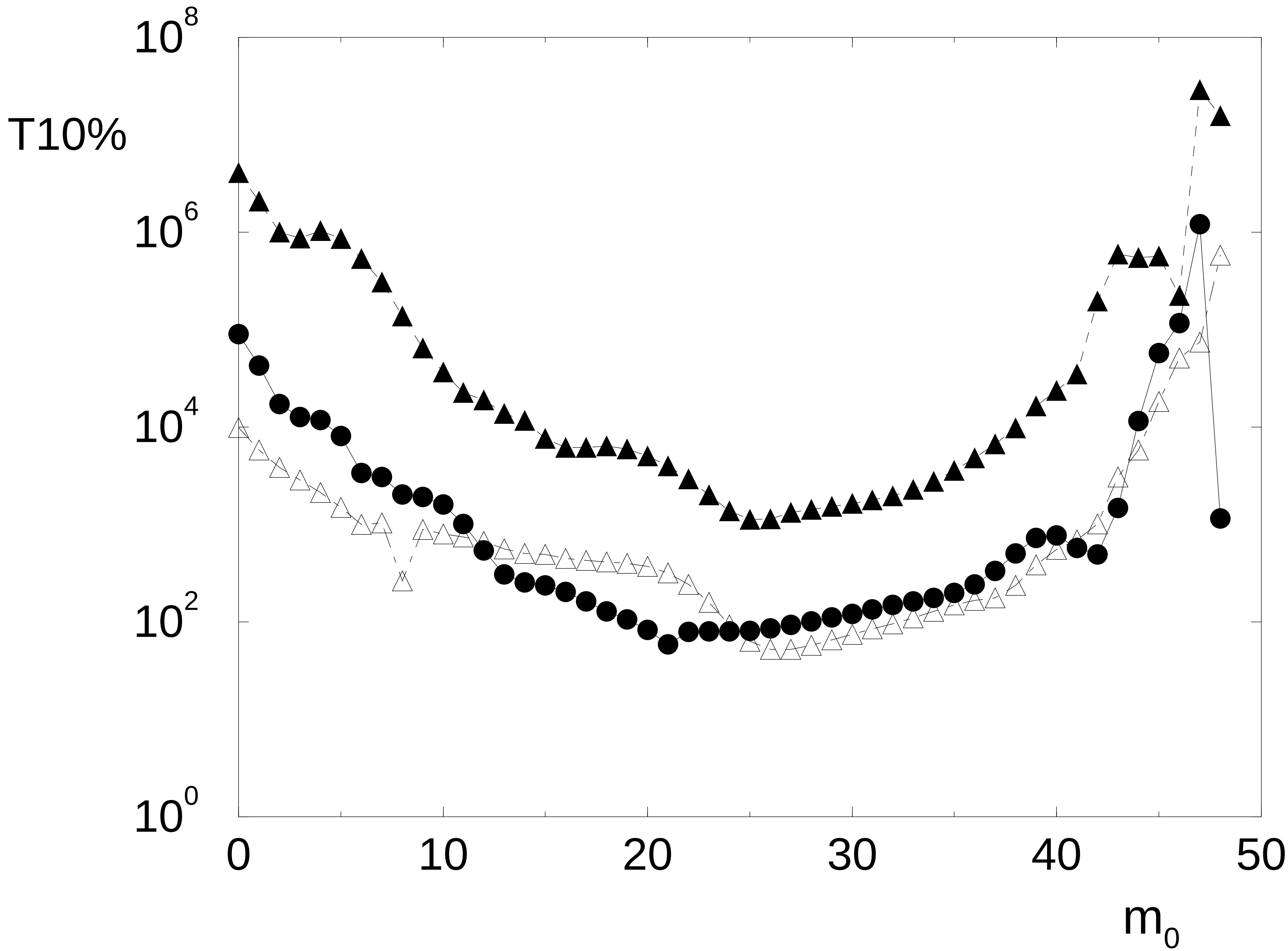

T10%
$10^8$
$10^6$
$10^4$
$10^2$
$10^0$
0
10
20
30
40
50
$m_0$

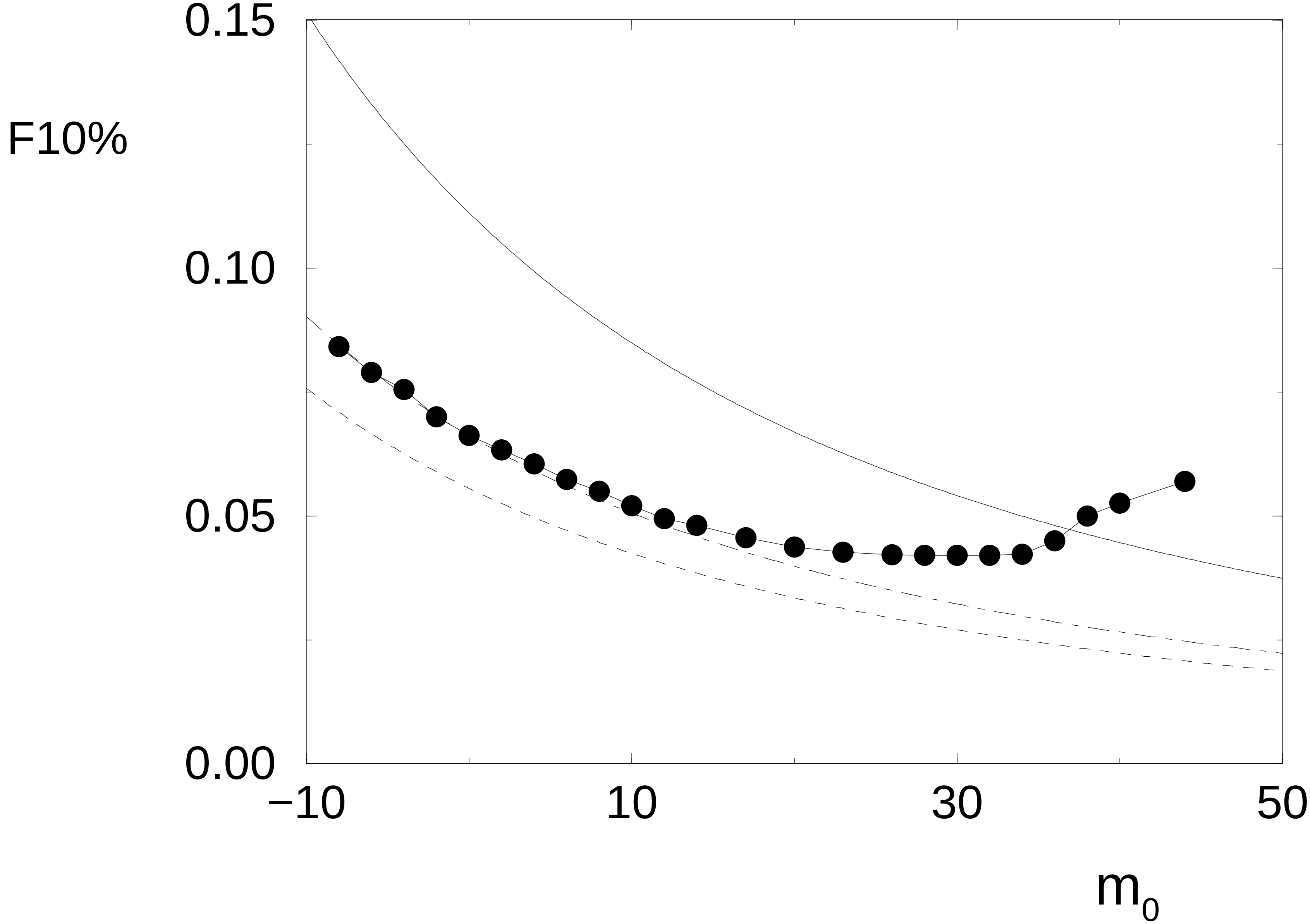

F10%
0.15
0.10
0.05
0.00
−10
10
30
50
m0

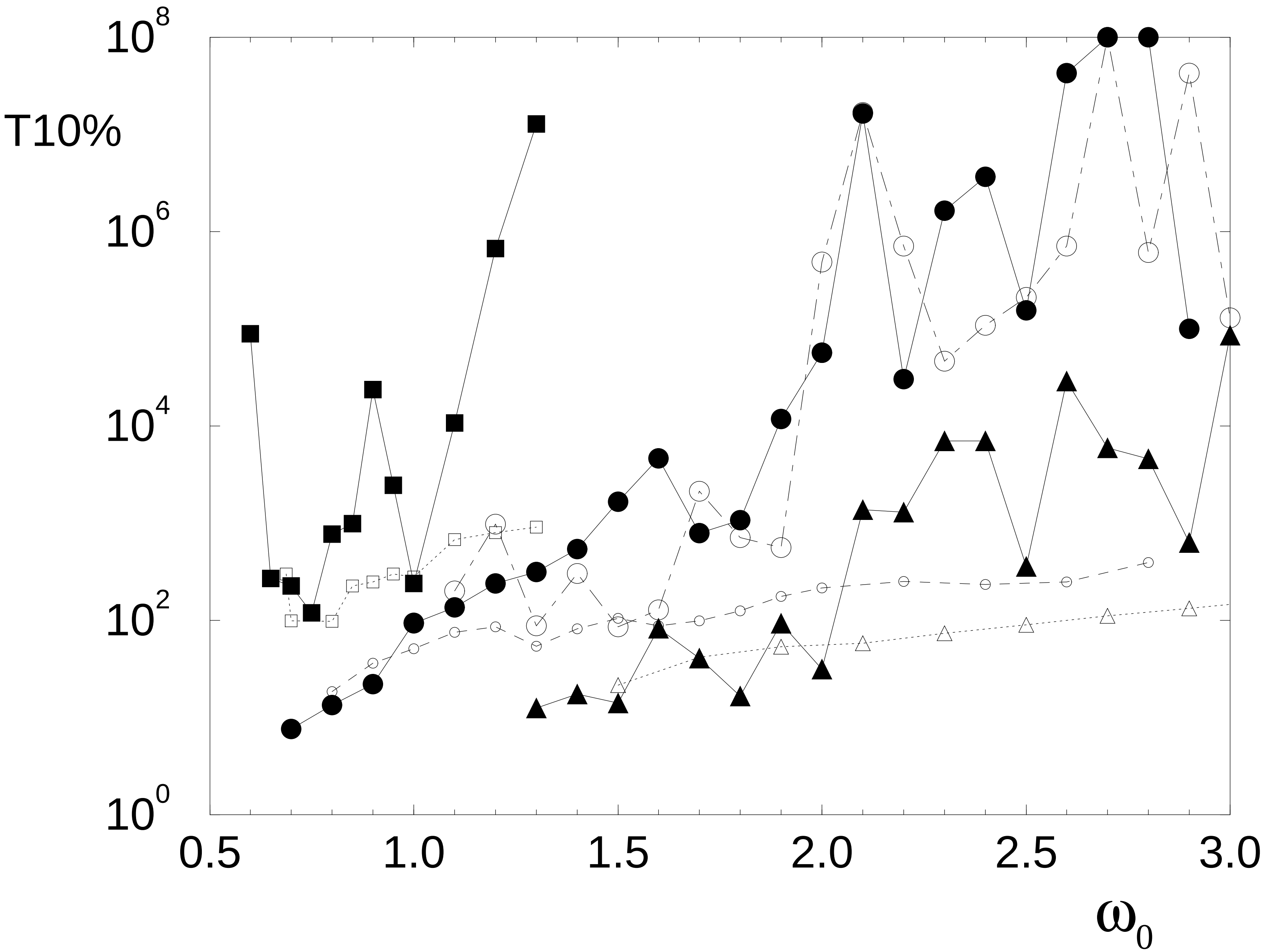
T10%
$10^8$
$10^6$
$10^4$
$10^2$
$10^0$
0.5
1.0
1.5
2.0
2.5
3.0
$\omega_0$

# Two-dimensional Quantum Hydrogen Atom in Circularly Polarized Microwaves: Global Properties

Jakub Zakrzewski[1,2], Robert Gębarowski[1*], and Dominique Delande[2]

[1] *Instytut Fizyki Mariana Smoluchowskiego, Uniwersytet Jagielloński, ul. Reymonta 4, 30-059 Kraków, Poland,*

[2] *Laboratoire Kastler–Brossel, Université Pierre et Marie Curie, T12, E1, 4 place Jussieu, 75272 Paris Cedex 05, France*

## Abstract

The ionization of hydrogen Rydberg atoms by *circularly* polarized microwaves is studied quantum mechanically in a model two dimensional atom. We apply a combination of a transformation to the coordinate frame rotating with the field, with complex rotation approach and representation of atomic subspace in Sturmian-type basis. The diagonalization of resulting matrices allows us to treat exactly the ionization of atoms initially prepared in highly excited Rydberg states of principal quantum number $n_0 \approx 60$. Similarities and differences between ionization by circularly and linearly polarized microwaves are discussed with a particular emphasis on the localization phenomenon. The dependence of the ionization character on the type of the initial state (circular, elliptical or low angular momentum state) as well as on the helicity of the polarization is discussed in detail.

*present address: Department of Applied Mathematics and Theoretical Physics, The Queen's University, Belfast BT7 1NN, Northern Ireland.

42.50.Hz,05.45.+b,34.50.Gb,32.80.Rm

Typeset using REVTEX

## I. INTRODUCTION

The hydrogen atom placed in an external field plays an exceptional rôle in the studies of quantum – classical correspondence in the vast area of quantum chaology. This system belongs to a small class of problems in this area where accurate theoretical predictions may be confronted with detailed experimental studies. This unique opportunity has led to a great progress in understanding of the behaviour of quantally chaotic systems for which both the experiments and the theory have been providing new ideas and new challenges. Moreover, the comprehension of the hydrogen atom interacting with strong fields is a prerequisite in studies of other atomic systems.

Despite over 20 years of intensive investigations [1], the theory of a highly excited hydrogen atom in a presence of a static uniform magnetic field still brings us unexpected predictions [2]. The ionization of highly excited hydrogen atoms by linearly polarized microwaves (LPM) has also a long history which began with the pioneering experiment of Bayfield and Koch [3]. As in the previous example, a complete physical picture of the coupled atom–field dynamics in this problem has yet to be reached. The very first model of the ionization process has been launched [4] using Monte–Carlo classical simulations, in which the ionization threshold was associated with the onset of classical chaos in the system. Numerous studies performed since this early work treated the problem either classically or quantum mechanically, at various degrees of approximation. At the same time, new improved experiments provided a stimulus as well as new puzzles for the theory (for recent reviews of the theory see [5–9], experimental details may be found in [10–12]).

A typical quantity measured or calculated in the ionization problem, is a microwave field amplitude required to produce 10% ionization yield as a function of the field frequency for a given duration time of the microwave pulse. Such a definition allows for a rough separation of the microwave frequency domain into a few regions, corresponding to different ratios of the microwave frequency, $\omega$, to the Kepler frequency on the initial orbit, $\omega_K$. For the scaled frequency, $\omega_0 = \omega/\omega_K << 1$, the classical ionization is due to an over the barrier escape (like for a static, homogeneous electric field case – the situation realized here in the $\omega_0 \to 0$ limit). Furthermore, in this frequency domain, quantum corrections due to the tunneling may be taken into account by means of semiclassical methods [13]. For higher frequencies (but still for $\omega_0 < 1$) the quantum ionization threshold is approximated quite well by the onset of classical chaos and the break–up of KAM tori. The diffusive gain of energy by the electron is the main mechanism leading to the ionization with some additional modifications due to classical resonances. For frequencies $\omega_0 > 1$, the physics of the ionization is quite different, because the quantal thresholds are significantly higher than the estimations of the classical model. This is attributed to the phenomenon of quantum photonic localization, which is analogous to the Anderson localization in disordered solids. Not surprisingly, it is this frequency domain which has been most intensively studied both experimentally and theoretically in recent years [5–12]. Needless to say, the discrepancy between classical and quantum predictions of the threshold for $\omega_0 > 1$ if of importance for a deeper understanding of the semiclassical limit. Finally, one envisions a "high frequency" domain, where a typical multiphoton ionization occurs. Certainly, in this regime, not only $\omega_0$ but also the (purely quantum) number of photons needed to reach the threshold, $\mathcal{N}_f$, becomes an important parameter characterizing the system.

The physics of hydrogen atom ionization by *circularly* polarized microwaves (CPM) is much less understood although first theoretical studies came about 15 years ago [14,15]. Casati, Guarneri and Shepelyansky, while formulating their photonic localization theory [6], considered the general case of elliptical polarization for a two-dimensional atom. The recent sudden growth of the interest in CPM ionization is certainly stinulated by the experiments carried out for alkali atoms [16,17] and new planned efforts for the hydrogen atom [18]. A number of classical studies has been reported [19–29]. In most of these works (with important exceptions of [21,27,29,30]) a simplified two-dimensional model of the atom has been used, i.e., the electronic motion plane coincided with the polarization plane. Also the only quantum study [31] considered such a simplified atomic model.

One reason for taking this simplification is purely technical. There is a basic difference between ionization by the LPM and the CPM. In the former case, the projection of the angular momentum on the polarization axis is conserved, thus the considered problem is effectively a two–dimensional one. On the contrary, the case of the CPM ionization requires, in principle, to study a fully three–dimensional (3D) system, because no constant of motion is known to exist apart from an approximate one in the purely perturbative regime [32].

Another reason for considering the simplified model emerges from LPM ionization studies. Namely, it turned out that a significant insight to the ionization may be obtained from the simplest, one–dimensional (1D) atomic model. Such a model may be quite well applicable for initial states extended along the polarization axis [6–9]. Such states are most vulnerable to microwave perturbation and determine the ionization threshold even for unpolarized initial states distribution (microcanonical ensemble). For the CPM ionization problem, as mentioned above, the simplest model is a two–dimensional one. As intuition immediately suggests, orbits lying initially in the polarization plane are most vulnerable to the microwave field and ionize at smaller microwave amplitude than inclined orbits [29]. Thus, as in the LPM, the simplified, lower dimensional model may be quite useful for estimating ionization thresholds.

Following the experiments [16,17], theoretical studies of the CPM ionization concentrated mostly on the low-frequency situation, $\omega_0 < 0.6$, both classically [19,21,23–25,27] and quantum mechanically [31]. The motion in this regime is mainly regular [24]. Different interpretations concerning the origin of the obtained ionization thresholds have been proposed. Since we shall consider only much higher frequencies here, we refer the reader to original works for the details.

Much less is known about the ionization mechanism in a higher frequency domain, $\omega_0 > 0.6$. For high eccentricity states and high frequencies ($\omega_0 >> 1$) the Kepler map has been constructed [6] yielding classically diffusive motion, as well as predictions for ionization thresholds, based, in the quantum case, on the photonic localization theory. Soon, however, Nauenberg has shown [20] that the original map of [6] is noncanonical. He proposed a new canonical map valid also for low eccentricity initial orbits. His results suggest that the electron motion, at least for circular or low eccentricity initial states, is quite regular. It is not inconsistent with the results of Chirikov resonance overlap analysis for the problem [22], where diffusive character of ionization has been pointed out for high eccentricity initial orbits, whereas, in the limit of vanishing eccentricity (circular states), the field amplitude given by the resonance overlap criterion diverges. A detailed numerical study, performed by us recently [26,29], supports in general the diffusive character of ionization for orbits of any eccentricity, stressing however, that the details of the ionization process are strongly dependent on the initial atomic state. Such a finding clearly indicates that the total phase space of the problem (at a given field value) is of a mixed type, with coexisting regular and chaotic domains.

The aim of the present paper is a *quantum* study of the CPM ionization of the hydrogen atom for the intermediate and the high frequency domain in the 2D model. This work extends our previous analysis [31] of low frequency ionization to a new domain, more interesting from the quantum chaos point of view. The exact quantum results are compared with the predictions of classical simulations for the 2D atom, obtained as in Ref. [29] with additional necessary modifications (see below). We also compare the quantum results with predictions of photonic localization theory [6]. As far as we know this theory has not been extended to the Nauenberg canonical map and no comparison of our quantum results with the latter quantized map is consequently possible.

The paper is organized as follows. The quantum treatment of ionization problem for the CPM is discussed in Section II. It utilizes the convenient property of the coupling between the atom and the circularly polarized wave, namely the possibility of removing the explicit periodic time dependence by a unitary transformation to the frame rotating with the field [36,37]. We show that for the CPM, this approach is equivalent to an application of the Floquet theory [35]. Section III contains the description of methods used to calculate the ionization probability and the definition of the ionization threshold. Due to the ambiguities of the commonly accepted microwave amplitude threshold for a fixed pulse duration, we define (in Section IV) a novel measure — a pulse duration threshold for a given microwave amplitude which we call the time threshold. In Section V the main numerical results concerning the ionization thresholds for various initial states are presented and compared with the classically obtained thresholds as well as with the photonic localization theory predictions. We discuss the origin of quantum-classical differences and we show the importance of the time scale over which the appreciable ionization takes place. We find also several classically scaling structures in the frequency dependence of the thresholds (section VI). Finally section VII contains the summary of the obtained results and the conclusions.

## II. THE THEORETICAL APPROACH

Let us consider first the two-dimensional model of hydrogen without any external perturbation. Its Hamiltonian (in atomic units) is obtained from the standard three-dimensional case by suppressing the $z$-dependence:

$$H_0 = \frac{p_x^2 + p_y^2}{2} - \frac{1}{\sqrt{x^2+y^2}}. \tag{2.1}$$

The quantum energies follow the Rydberg formula, $E_n = -1/2n_*^2$, where the effective principal quantum number, $n_* = n + 1/2,\ n = 0, 1, 2, ...$, appears rather than $n$ itself. The states may be characterized by two quantum numbers: $(n, m)$, where $m = -n, .., n$ is the eigenvalue of the angular momentum operator $L_z = xp_y - yp_x$.

The atom in the field of the circularly polarized radiation is described by the Hamiltonian

$$H = \frac{(\vec{p} + \vec{A})^2}{2} - \frac{1}{r} \tag{2.2}$$

with the vector potential, $\vec{A}$,

$$\vec{A}(t) = \frac{F(t)}{\omega}\left(-\vec{e_x}\sin(\omega t) + \vec{e_y}\cos(\omega t)\right). \tag{2.3}$$

The electric field reads $\vec{\mathcal{F}} = -\partial\vec{A}/\partial t = F(t)\left(\vec{e_x}\cos(\omega t) + \vec{e_y}\sin(\omega t)\right)$ provided $F(t)$ changes slowly with respect to $\omega^{-1}$; $dF/dt << \omega F$. Hence $F(t)$ is simply the amplitude of the electric field. In the following we will mainly consider an interaction of atoms with the field of a constant amplitude, $F(t) = F$.

Note that a change of the sign of $\omega$ in Eq. (2.3) is equivalent to the change from the right-hand CPM to the left-hand CPM. We shall explore this possibility below and allow $\omega$ to take both positive and negative values in order to study both types of polarization.

Let us begin with a constant $F$ problem. Expanding the kinetic energy in Eq. (2.2) and removing the constant $A^2$ term, one arrives at the Schrödinger equation of the form

$$i\frac{\partial}{\partial t}|\psi>= \left(\frac{p^2}{2}-\frac{1}{r}-\frac{F}{\omega}\left(p_x\sin(\omega t)-p_y\cos(\omega t)\right)\right)|\psi> . \tag{2.4}$$

The Hamiltonian in Eq. (2.4) is an oscillatory function of time. Therefore, one may use the Floquet theory [35] frequently applied to ionization problems both in the optical and microwave regimes. We adopt here another approach, verified already in our study of low frequency domain [31]. Namely, the time dependence for the circular polarization is removed by passing to the rotating frame [36–38]. Under the unitary transformation $U=\exp(i\omega L_z t)$, Eq. (2.4) becomes

$$i\frac{\partial}{\partial t}|\tilde{\psi}>= \left(\frac{p^2}{2}-\frac{1}{r}+\frac{F}{\omega}p_y-\omega L_z\right)|\tilde{\psi}>, \tag{2.5}$$

where $|\tilde{\psi}>=U|\psi>$. Since the Hamiltonian is now time independent, we may look for the eigenvalues and eigenvectors of the corresponding time–independent Schödinger equation. In the following we pass to the scaled semiparabolic coordinates, defined as $u=\sqrt{(r+x)/\Lambda}$, $v=\sqrt{(r-x)/\Lambda}$, already used successfully in the treatment of the hydrogen atom in a static magnetic field [1]. These coordinates allow to remove the Coulomb singularity and facilitate accurate classical simulations of the problem [21,26,29]. The Schödinger equation takes the form of the generalized eigenvalue problem

$$\left(\frac{p_u^2+p_v^2}{2\Lambda^2}-\frac{2}{\Lambda}-\frac{F}{\omega\Lambda}(vp_u+up_v)-\frac{\omega}{2}(u^2+v^2)(up_v-vp_u)\right)|\tilde{\psi}>=E(u^2+v^2)|\tilde{\psi}> . \tag{2.6}$$

All terms in the equation above have the form of a polynomial in coordinates and momenta. This form suggests the use of the harmonic oscillator basis for an efficient diagonalization of the problem. The scale parameter, $\Lambda$, may be chosen at will and is related to the common frequency of oscillators in $u$ and $v$ coordinates. Introducing standard creation and annihilation operators,

$$u = \frac{1}{\sqrt{2}}(a^\dagger + a), \quad v = \frac{1}{\sqrt{2}}(b^\dagger + b) \tag{2.7}$$

$$p_u = \frac{1}{i\sqrt{2}}(a^\dagger - a), \quad p_v = \frac{1}{i\sqrt{2}}(b^\dagger - b) \tag{2.8}$$

allows for a simple calculation of matrix elements in Eq. (2.6). However, the matrix representation of the $L_z = (up_v - vp_u)/2$ operator becomes complex. Therefore, it is more convenient to introduce circular creation and annihilation operators for atomic variables which obey the circular symmetry invoked by the field,

$$A^\dagger = \frac{1}{\sqrt{2}}(a^\dagger + ib^\dagger), \quad A = \frac{1}{\sqrt{2}}(a - ib) \tag{2.9}$$

$$B^\dagger = \frac{1}{\sqrt{2}}(a^\dagger - ib^\dagger), \quad B = \frac{1}{\sqrt{2}}(a + ib) \tag{2.10}$$

Expressed in terms of these operators, $L_z$ reads

$$L_z = \frac{1}{2}(up_v - vp_u) = \frac{1}{2}i(ab^\dagger - a^\dagger b) = \frac{1}{2}(A^\dagger A - B^\dagger B) \tag{2.11}$$

and is diagonal in the circular oscillator basis, $|n_A, n_B>$. The meaning of the scale parameter, $\Lambda$, is clear now. By choosing $\Lambda = n$ in the absence of the microwave field, one notices that the state $|n_A, n_B>$ is an eigenvector of the Hamiltonian corresponding to the energy $E = -1/2(n+1/2)^2 - m\omega$, where $m = \frac{1}{2}(n_A - n_B)$ and $n = \frac{1}{2}(n_A + n_B)$ [43]. Therefore, this is the eigenstate of the field–free 2D hydrogen in the rotating frame. The oscillator basis is equivalent to the so called Sturmian basis in the original Cartesian coordinates with $\Lambda$ being the parameter of the Sturmian basis (for the detailed discussion, see [39]). The basis set, for arbitrary $\Lambda$, will be denoted as $\{|n, m>_\Lambda\}$.

In the chosen basis $\{|n, m>_\Lambda\}$ the relation of the eigenvalues, $E_i$, and eigenvectors, $|\tilde{\psi}_i> = \sum_{n,m} a_{nm}|n, m>_\Lambda$, of Eq. (2.6) to Floquet eigenstates becomes transparent. In the laboratory frame the $i$–eigenvector becomes

$$|\psi_i(t)> = \exp\{-iL_z\omega t\}\,|\tilde{\psi}_i> = \sum_{n,m} a_{nm} \exp\{-im\omega t\}\,|n, m>_\Lambda\ . \tag{2.12}$$

Indeed, Eq. (2.12) shows a typical periodic time behaviour characteristic for a Floquet eigenstate. It may be verified after straightforward algebra [40] that, starting from the usual

Floquet Hamiltonian [35] corresponding to Eq. (2.4) and passing to the similar oscillator representation, one obtains the Floquet matrix in the oscillator–Floquet basis $\{|n, m>_\Lambda |K>\}$ ($K$ enumerates the photon blocks), which has a block diagonal structure due to the selection rule, $m + K = M = const$. Eigenvalues of different $M, M'$ blocks differ by the value $(M - M')\omega$, as it is in a typical, periodic in $\omega$ Floquet structure. Therefore, the diagonalization of Eq. (2.6) is equivalent to finding all independent eigenvalues and eigenvectors of the Floquet Hamiltonian. Note, however that the above mentioned selection rule appears only in the circular polarization and it is a direct manifestation of the well known from atomic physics, $\Delta m = 1$ selection rule for the absorption of a circularly polarized radiation. This rule is still valid for our Sturmian basis $\{|n, m>_\Lambda\}$ since its angular properties are the same as that of atomic wavefunctions diagonal in angular momentum representation.

In the presence of electromagnetic field, the system, Eq. (2.6), supports no bound states but rather resonances, due to the coupling to the continuum. To find these resonances and the corresponding wavefunctions, we use the well known complex rotation technique [45]. In short, one rotates the coordinates and momenta according to $\vec{r} \to \vec{r}\exp(i\theta), \vec{p} \to \vec{p}\exp(-i\theta)$. The matrix representing the Hamiltonian in Eq. (2.6) becomes then complex symmetric and explicitly dependent on the rotation angle $\theta$. The rotation does not affect energy values of the bound states while rotating structureless continua by the angle $2\theta$ in the complex plane. Provided the rotation angle is not too small, the resonances in the continuum become "exposed" by the rotation and appear as complex eigenvalues $\epsilon_i = E_i - i\Gamma_i/2$. For sufficiently large basis set, an interval of $\theta$ values exists for which positions of resonances in the complex plane are independent of the rotation angle and the scale parameter $\Lambda$. Consequently, the real part of $\epsilon_i$ is interpreted as the resonance position and $\Gamma_i$ is the width of the resonance (an inverse of its lifetime). Moreover, the corresponding wavefunctions are square integrable.

The diagonalization of Eq. (2.6) yields *exact* (within numerical errors) eigenvalues and eigenvectors of the hydrogen atom coupled to the monochromatic electromagnetic field – i.e., we find dressed atom + field states (Floquet eigenstates) for a given value of the microwave amplitude $F$. This approach to treat nonperturbatively the ionization of hydrogen atoms driven by a periodic wave is fairly standard in a sense that it combines the complex rotation technique with the discrete basis set (Sturmian) representation in order to include correctly the coupling to the continuum. Such an approach, originating as far as we could trace from the work of Chu and Reinhardt [46], has been used intensively for studies of various aspects of the strong field ionization of the hydrogen atom for both the linearly and circularly polarized radiations [47,48]. However, these studies were mostly related to the ionization from low lying atomic initial states. In the microwave regime, the same technique was adopted for the LPM by Buchleitner and Delande [9,49,50] for both 1D and fully 3D situation (effectively 2D due to conservation of $L_z$ — see the introduction). Our contribution to the method includes its adaptation for the 2D hydrogen atom in the CPM, and also the application of a simple algebra of harmonic oscillators, which allows for an efficient and straightforward matrix evaluation.

## III. IONIZATION PROBABILITY

In the case when the amplitude, $F = F(t)$, changes in time, results obtained for various $F$ values correspond to instantaneous dressed states. Clearly, finding the Floquet states does not provide the exact solution of the full time-dependent Schrödinger equation for arbitrary microwave pulse, $F(t)$. Two possible ways of generating an approximate solution out of the results obtained for diagonalizations of the Floquet matrix are quite commonly used.

First, let us assume that the atom is in a state $|\varphi_0>$ before the microwave radiation is turned on. For sufficiently slowly varying microwave pulse $F(t)$, the atom has enough time to adjust to the changes in field amplitude – this is nothing else but the adiabatic approximation. Under this approximation, it is sufficient to follow the single dressed (Floquet) state which is a smooth continuation of the initial state, $|\varphi_0>$. Such an approach will be referred to as a single Floquet state approximation (SFSA). The energy of the state and its width change with $F(t)$, providing in this way the information about the nonresonant decay of the atomic population to the continuum. Therefore, the probability of ionization after the microwave pulse of a duration $T$ is given simply by

$$P = 1 - \exp\left\{ - \int_0^T \Gamma[F(t)]\, dt \right\}, \tag{3.1}$$

where $\Gamma$ is the width of the important single Floquet state.

In the presence of quantum resonances (between field shifted states) a real population transfer to other states may occur and the SFSA becomes questionable. Such resonances appear as avoided crossings between energy levels as $F(t)$ changes in time. It is well known from the Landau-Zener theory that even in the presence of avoided crossings, the single state approximation may work in two extreme cases, either for a very slow adiabatic passage or for a fast diabatic passage. The characteristic time scale is set by the inverse size of the avoided crossing gap (minimal distance between levels), $\tau = 1/\Delta\epsilon$. If a change in $F$, sufficient to pass the avoided crossing, occurs in time $\Delta t$ then the adiabatic passage is realized for $\tau \ll \Delta t$. The diabatic transition occurs in the opposite limiting case, i.e. for very narrow avoided crossings.

It is now apparent that the relative sizes of avoided crossings encountered during changes of $F(t)$ are very important. For classically regular systems, the avoided crossings are generically much smaller in size than the mean level spacing. Thus, in the semiclassical limit, one expects that the SFSA can work for the classically regular regime in the situation where avoided crossings are passed diabatically, while in between subsequent avoided crossings, the single level is adiabatically followed (as the size of a typical avoided crossing is much smaller than the mean level spacing). Such a mixed adiabatic-diabatic picture has been commonly assumed for the ionization from low lying states [47,48]. Here we relate it to the character of the classical dynamics in the semiclassical limit appropriate for the microwave ionization from highly excited initial states. On the other hand, for classically chaotic systems, the avoided crossings of arbitrary size appear abundantly in a typical spectrum [53]. Thus, one expects a breakdown of the SFSA when the ionization is classically chaotic. However, the SFSA may still work quite well in the mixed phase space for some of the states, either those located in the regular regions, or even those strongly localized on classical phase–space struc-tures such as periodic orbits, cantori, separatrices etc. Such states, due to their localization properties, are generically involved in avoided crossings narrower than the average.

Typical microwave pulses used in experiments [10–12,51] have a time duration ranging from a few hundred up to thousands of field cycles. For such long times, a direct integration of the time–dependent Schrödinger equation is not feasible. In most of the experimental setups, the microwave pulse has the "flat-top" shape (with turn-on phase, constant $F$ amplitude dominant phase and the pulse turn off). Hence it is reasonable to model this situation by a rectangular pulse of a constant amplitude $F$ during the pulse duration $T$ [9]. In this rectangular pulse approximation (RPA), a single diagonalization is enough to obtain the ionization probability as

$$P = 1 - \sum_i \mathcal{R}(< \tilde{\psi}_i|\varphi_0 >^2)\exp(-\Gamma_i T), \tag{3.2}$$

where the sum is over dressed states. The scalar product should be calculated without complex conjugation (as appropriate for complex symmetric matrices) and $\mathcal{R}$ stands for the real part. As discussed in detail by Buchleitner [9,50], this expression is exact in the limit of long microwave pulses after averaging over the initial phase of the microwave field.

It is questionable whether the RPA, Eq. (3.2), may provide us with the information about the real physical process when the pulse changes smoothly in time. Such a situation is expected in particular for a strongly chaotic system when the classical ionization is mainly diffusive. In that case, one expects that the actual wavefunction, e.g., in the "flat-top" region of the pulse, is a linear combination of many Floquet states, i.e., its span on the dressed eigenbasis is large [54]. On the other hand, the more regular behaviour of the system, the worse the response of the system to the real pulse will resemble the one obtained from the RPA (and the better the SFSA will be as discussed above). The two described methods are somehow opposite in nature, and become equivalent only in the perturbative limit, when a single overlap in Eq. (3.2) practically exhausts the sum.

As shown in the classical study [29], except in the regimes of very low and very high frequencies (with respect to the Kepler frequency of the initial orbit), the classical ionization has a diffusive, chaotic character. Therefore, most of the results shown in the next sections use the RPA and Eq. (3.2) to calculate the ionization probability. The aim of such an approach is to show a global frequency dependence of the ionization threshold, while being aware of all the limitations of the RPA as far as the comparison with experiments [10–12,51] is concerned.

## IV. THE IONIZATION THRESHOLD

Experimental results are typically plotted [10–12] using the so called scaled variables. The origin of scaling for an atom placed in an external field is purely classical [1]. As a result, one uses scaled frequency, $\omega_0 = \omega n_*^3$ (i.e. the ratio of the microwave frequency to the Kepler frequency $\omega_K$ of the initial orbit at energy $E_0 = -1/2n_*^2$) and the scaled microwave amplitude $F_0 = F n_*^4$ to characterize the ionization process. Similarly, pulse duration may be measured in Kepler periods or directly in the number of cycles of the microwave field. Therefore, results represented in the scaled variables allow for a direct comparison of data obtained for initial states with different principal quantum number. Since the scaling is not preserved quantum mechanically, the experimental and theoretical results for the ionization yield depend on the initial state for the same scaled parameters – this difference should, however, vanish in the classical limit.

The commonly accepted definition of the ionization threshold is the so called amplitude 10% threshold, $F10\%$, i.e. the value of the microwave amplitude for which 10% of atoms become ionized. The dependence of the $F10\%$ threshold on scaled frequency $\omega_0$ proved to be a useful characteristic [10]. Recently, such curves showing changes of the $F10\%$ versus $\omega_0$ have been extensively studied classically for the CPM ionization in [29] for various initial orbits characterized by their eccentricities. In particular, we have discussed in detail the thresholds for orbits corresponding to circular and highly elliptical states lying in the CPM polarization plane. For such states, the simplified 2D quantum atomic model should be at its best while the corresponding classical model is exactly 2D. It is also now feasible to prepare such states in the laboratory [33,34].

It has been shown [29] that the ionization is prohibited classically for a sufficiently large mismatch between the frequency of the driving field and the Kepler frequency for low eccentricity states (a similar effect has been found earlier for the linear polarization [42]). It can be observed as one order of magnitude increase of the ionization threshold, which occurs around frequency $\omega > 3\omega_K$ and $\omega < -\omega_K$ for the initial circular orbits. This symmetry around $\omega = \omega_K$ rather than $\omega = 0$ is easily understood — the electron moves around the nucleus with frequency $\omega_K$, and it is the relative frequency of the electronic motion and the field which is important for the interaction rather than the field frequency alone. It is interesting to investigate whether the similar prohibition of high frequency ionization occurs in the quantal world. The results for the initial circular state $|n_0 = 24, m_0 = 24>$ are compared with the corresponding classical threshold in Fig. 1. Since the classical results have been obtained for a smooth pulse of rise/turn off time $T_r = 25$ Kepler periods and total duration $T = 500$ in the same units, the quantum curve has been obtained using the SFSA, Eq. (3.1), assuming the same pulse shape [44]. One may note that the increase of the quantum threshold also commences around frequency $\omega_0 \approx 3.2$, although it is much less rapid than the classical one and saturates at a value which is roughly twice higher than the (average) threshold for lower frequencies. Still the figure suggests that for $\omega_0 > 3.2$, the ionization has a purely quantum (most probably tunneling) origin. The single state approximation used in these calculations should be justified, because the system is classically regular at microwave amplitude values considered here.

It is worth to point out that the time of the pulse rise/turn off is just 10% of the total pulse length. In addition, as it is in the SFSA, the ionization rate increases with $F$, and consequently, the contribution of the pulse rise and the turn off to the total ionization probability is very small. Hence, results of the SFSA do not change in fact if we assume an almost rectangular (but smooth) pulse. On the other hand, Fig. 1 presents the thresholds found for the RPA, according to Eq. (3.2), which differ significantly from those obtained using the SFSA. In particular, the increase of the RPA threshold for higher frequencies is much less pronounced than for the SFSA. This feature can be easily explained – the microwave amplitude is large, $F10\%$ is of the order of 0.3 atomic unit, and we are beyond the perturbative limit. Indeed, although there is one dominant Floquet state in the expansion, Eq. (3.2), its overlap on the initial state is only approximately 0.6. It is this state which is followed in the SFSA, but in the RPA this state carries less significant contribution to the ionization. In the RPA, the 10% ionization is reached mainly due to the fact that this method leads to the excitation of other Floquet states which decay more rapidly. Clearly, the realistic physical situation is reproduced by the SFSA and these results are relevant for a comparison with a future experiment.

We show the results for the high frequency domain in Fig. 1. This is the region where the SFSA may work - it is not so for a lower frequency domain, where the classical ionization has a chaotic (diffusive) character [29]. Another, even more important, reason for the limitation on frequency interval is the fact that the $F10\%$ amplitude threshold becomes quite ambiguous for lower frequencies. These problems are exemplified in Fig. 2 where we present the ionization yield as a function of the microwave amplitude $F$ for rectangular pulses of different durations. The essential point is that for any pulse duration, the ionization probability is a *non-monotonic* function of $F_0$, thus several $F_0$ may yield the same ionization probability. As shown in the figure, the local maxima and minima occur for different probability values, so a simple change from 10% to, say, 50% ionization does not help. Note, parenthetically, that such a sharp structure is not typically observed in the experiment due to variations of the amplitude with respect to the cavity axis as well as other effects that smear out most of the structures. We have checked that the non–monotonic behaviour is not an artifact of the RPA used. A qualitatively similar behaviour is observed in the SFSA and may be linked to the influence of avoided crossings, as discussed in detail in [54,55].

From the theoretical point of view, it is advantageous to have an unambiguous definition of the threshold. It directly follows from the ionization probability formula, Eq. (3.2), that the ionization probability is a strictly monotonic function of pulse duration $T$ (the realistic values of the field amplitude are orders of magnitude smaller than those needed for a saturation of bound–free transitions, which may lead to Rabi–like oscillations and a non–monotonic time dependence). We propose, therefore, and use mostly hereafter, the novel definition of the 10% threshold defined as a pulse duration $T10\%$ (in microwave cycles) for which the 10% ionization probability is obtained for given values of the scaled frequency and the microwave amplitude.

The time threshold has additional advantages, both practical and of the fundamental interest. The latter, as we shall exemplify below, stems from the fact that the classical – quantum correspondence strongly depends on the pulse duration. This is an aspect of the problem which remained almost unexplored in most of the studies of the microwave ionization (see however Ref. [51]). In practical terms, the time 10% threshold is advantageous because it is much easier to find than the field 10% threshold. In order to determine the 10 % time threshold for a given initial state, frequency and microwave amplitude, it is sufficient to perform only a single diagonalization. This should be compared with several attempts at different $F_0$ values, required to estimate the amplitude 10% threshold. Since initial principal quantum numbers of the order of 50 or more are used experimentally, the corresponding matrices are quite large (see below) and a search for the amplitude threshold becomes very CPU–time expensive.

The example of the time threshold, $T10\%$, is presented in Fig. 3, for the same "testing" case of the circular state as in Figs. 1,2 for a rectangular pulse of amplitude $F_0 = 0.1$. Here, we compare also the SFSA to the exact results based on Eq. (3.2). Note a good agreement between both approaches for high frequencies (corresponding to a classically forbidden ionization, i.e., the classically regular dynamics). In contrast to that, there are much larger differences (sometimes orders of magnitude) between predictions of both methods for lower frequencies, where the classical motion is mainly chaotic at this value of $F_0$ (in agreement with the semiclassical picture discussed above). Note also that, apart from the resonance structures to be discussed in detail in following sections, longer and longer pulses are needed to obtain the 10% ionization yield as the microwave frequency is increased. This behaviour reflects the similar increase of the corresponding classical curve for the fixed pulse duration [29]. In fact the agreement between the RPA and the SFSA for high frequencies reflects the perturbative character of the ionization in this limit for $F_0 = 0.1$ (the overlap of the initial state on the most important Floquet state is larger than 0.92).

At this value of $F_0$, pulses shorter than 100 microwave cycles suffice to yield 10% ionization for $\omega_0 < 2$. On the other hand, for high frequencies, the required pulse duration may well exceed $10^6$ cycles. This is an illustration of the fact that, when the atomic response strongly varies with frequency, one should adjust the microwave amplitude to the studied frequency interval so as to remain within a reasonable, experimentally accessible, range of pulse durations. Still, finding $T10\%$ for a few values of the microwave amplitude is much more numerically efficient than attempting to determine precisely $F10\%$ for a fixed pulse duration.

## V. RESULTS OF NUMERICAL SIMULATIONS FOR A RECTANGULAR PULSE

The results presented up till now served to illustrate our approach and to test possible approximations. They prove that, in the chaotic regime we are interested in, the RPA is the most reasonable approximation. To simulate ionization of Rydberg states of principal quantum number $n \approx 60$ typically met in the experiments [10–12], the results for $n = 24$, shown above, are not truly convincing. In this section, we present $T10\%$ thresholds obtained for $n = 48$ and $n = 64$, and the comparison with the corresponding classical simulations. It is thus probably a good point now to discuss some technical details of the calculations. More detailed description can be found in an earlier analysis of the LPM ionization by the same approach [50].

The Sturmian basis $\{|n, m >_\Lambda,\ n = 0, 1, ...;\ m = -n, ..., n\}$ is complete, however, for practical purposes, we have to limit the size of the basis set. Typically we take the basis vectors up to some $n_{max}$. In the ideal case (and it was the case for the results presented up till now) $n_{max}$ should be chosen sufficiently big, so that any further increase would not affect the results. The value of the Sturmian parameter $\Lambda$ is chosen typically in such a way that one of the Sturmian functions coincides with the initial state studied.

Real experiments [10,11] do not measure the absolute ionization yield. Both the stray electric fields present in the experimental setup and the method of detection do not allow to differentiate between real ionization and excitation of the atom to a very high Rydberg state (with e.g., principal quantum number $N$=89 or $N$=114, depending on the experiment [11]). Thus, to simulate a genuine experimental situation, one should count such a high excitation case as "ionization". Fortunately, one may use properties of the Sturmian basis to perform such a selection in a very effective and only slightly non–rigorous way as shown in Ref. [50]. Up to a given $n_{max}$, the Sturmian basis faithfully represents only a finite region of the configuration space. Discrete states which extend further, when represented in the "too small" basis act like continuum states. The effect is clearly seen when diagonalizing the complex rotated atomic Hamiltonian in the absence of any perturbation. The relatively low lying states are well converged and lie (as they should) on the real axis, the continua are rotated (as they should) by the angle $2\theta$ [45], but the ionization threshold is shifted to smaller energies. It is easy to show (by looking at the Sturmian functions in the coordinate representation [50]) that $N \approx \sqrt{\Lambda n_{max}}$ is the effective principal quantum number at which the shifted threshold appears. Therefore, limiting the $n_{max}$ value, one may very efficiently realize experimental high–$N$ cutoff and at the same time limit the size of diagonalized matrices.

For the atom in a given initial state $|n_0, m_0>$, it is elementary to find out how many photons are required for the ionization in the perturbative picture. Since we are beyond the perturbative limit, it is necessary to include all basis states which may be reached by a few photons more than the minimal required number. Due to the $\Delta m = \pm 1$ selection rule, it is sufficient to form the matrix among the basis states accessible from the initial angular momentum $m_0$ value by, say up to $K$ photons, i.e., to restrict the basis states to the interval $m_0 \pm K$. Still the size of resulting matrices may well be up to 25000. By changing the rotation angle $\theta$ and the number of photons $K$, we were able to verify the convergence of our results.

In order to evaluate the ionization probability, Eq. (3.2), the eigenvalues and eigenvectors which have non negligible overlap with the initial state are needed. We used the Lanczos diagonalization method adapted for complex symmetric matrices [52], which allows to obtain a subset of eigenvalues and eigenvectors around a given value of energy. We started the process around the field–free rotated energy and continued the Lanczos scheme until the eigenvectors found exhausted the norm of the initial state up to 0.995. Typically few percent of all eigenvectors were sufficient to satisfy this criterion.

Finally let us briefly describe the classical calculations performed as in [29], but with two modifications. First, we have included in the calculations the "effective threshold" criterion for the ionization to enable comparison with quantum data. Thus we counted as "ionized" any trajectory with final (unscaled) energy greater than $-1/2N^2$. Second, we have also determined classically $T10\%$ instead of $F10\%$ thresholds. The classical results were obtained obviously for a rectangular, not smooth pulse and for the 2D model of an atom.

As the problem studied is two–dimensional, the quantum dynamics may be (and it is in fact) strongly dependent on the geometrical properties of the initial atomic state (in agreement with classical mechanics, similarly sensitive [29]). We cannot discuss all the possible initial states for obvious reasons and we have to restrict ourselves to some illustrative examples. Hence to introduce some systematics into the presentation, first we show results for the elongated states (small $m_0$) using as an example $m_0 = 0$ state behaviour. Then we consider states of "average" ellipticity ($m_0 = n_0/2$), and finally, the circular states ($m_0 = n_0$).

### A. Elongated states

Let us consider first the minimal angular momentum states, $m_0 = 0$. Such states are extremely important for the LPM ionization, where they are most vulnerable for perturbation and among states of different angular momenta, they ionize at lowest amplitudes of the microwave pulse [9,49,50]. For such states, a simplified 1D model can be used, in which the motion of the electron is restricted to the field axis. The behaviour of this model has been studied in great detail for the LPM ionization [5–7]. In particular, a lot of attention has been paid to the differences between classical predictions and quantum results, the differences which are commonly interpreted as a manifestation of the quantum localization phenomenon, being the "quantum chaology" analog of the Anderson localization in disordered solids. It is interesting to see whether similar differences are present also for the CPM ionization case.

Fig. 4 shows the time $T10\%$ thresholds found for moderate frequencies for different values of the microwave amplitude $F_0$ in comparison with the corresponding classical simulations. For low frequencies, quantum results decrease with the frequency (as seen for the smallest amplitude data) — this is a signature of the small frequency behaviour, studied earlier [23,31], where the motion becomes regular (see discussion in [31]).

For a given scaled field value, the results are presented in the interval of $\omega_0$ where quantum 10% ionization occurs for times between 10 and say $10^8$ field cycles. The upper limit is sufficiently high to make plausible experimental verification of our calculations. On the other hand, the resonance widths corresponding to such long interaction times are quite small. They are in fact on the border of the numerical accuracy in the double precision arithmetics. The lines crossing the upper border of the figure indicate the frequency regime in which we could not, for that reason, determine accurately the widths. Since the corresponding $T10\%$ are outside the experimental range, they are not of major interest: in such a regime, the atom will be observed as stable against the ionization.

Note that, with increasing $F_0$, the curves typically shift to higher frequencies indicating that the ionization yield decreases with $\omega_0$ for fixed values of the microwave pulse amplitude and its duration. By comparison, the corresponding classical time thresholds (shown as dotted lines) are almost insensitive to the frequency change in the presented frequency interval shown. In effect, the agreement between the classical and quantum predictions is strongly dependent on the frequency but also, importantly, on the time needed for reaching the threshold. The classical and quantum threshold agree quite well, especially for low frequencies ($\omega_0 < 1$), when 10% ionization is reached in less than one hundred microwave cycles. At higher frequencies, classical and quantum results differ by orders of magnitude in values of $T10\%$.

These results may be confronted with the prediction of the localization theory as based on the so-called Kepler map [6]. The applicability of the Kepler map for the 2D CPM ionization case is limited to the regime:

$$\omega_0 \gg 1 \qquad |\tilde{m}| = \frac{|m_0|}{n_0} < \left(\frac{2}{\omega_0}\right)^{1/3}, \tag{5.1}$$

where $\tilde{m} = m_0/n_0$ is the scaled angular momentum taking values between -1 and 1 [6]. Clearly, the second condition may be important for elliptical states, discussed later. For the present case of $m_0 = 0$ only the frequency condition is relevant.

If the photonic localization takes place, then the maximal excitation, up to exponential corrections, is given by the localization length (in number of photons) [6]

$$\mathcal{L} \approx 3.33 n_0^2 \mathcal{H}^2 F_0^2 \omega_0^{-10/3} \tag{5.2}$$

where

$$\mathcal{H} = 1 + \frac{\tilde{m}^2}{2} + 1.09\omega_0^{1/3}\tilde{m}, \tag{5.3}$$

(see Eq. (63) and Eq. (71) of [6] with appropriate identification of symbols; in particular notice the difference of the factor $\sqrt{2}$ in Eq. (5.3) due to the similar difference in the definition of the microwave amplitude). Now let us assume that $F_0$ is sufficiently small to limit the excitation to the region below the effective ionization threshold. Hence, due to the photonic localization, the corresponding $T10\%$ is infinite: for such small fields, 10% ionization is never reached. The increase of $F_0$ will lead to the decrease of $T10\%$ through all orders of magnitude, and finally the other extreme limit is reached, the fast ionization limit, where a significant ionization occurs faster than the time necessary for the localization to build up.

This variation of $T10\%$ is thus a sensitive test of the relevance of the localization theory. Curves representing time thresholds over all possible time scales are plotted in Fig. 4. As before, the sharp growth of $T10\%$ with frequency and its shift towards higher frequencies with increasing $F_0$ is in a qualitative agreement with the photonic localization theory [6] and the corresponding decrease of the localization length, Eq. (5.2). It is worth noting that, for the smallest amplitude $F_0 = 0.025$, presented in Fig.4, the increase in $T10\%$ due to "localization" occurs already for values of $\omega_0$ slightly smaller than unity.

In a typical laboratory measurement or numerical experiment, the variations of the $F10\%$ threshold are quite small, they change at most by a factor 2. On the contrary, the $T10\%$ threshold can cover easily a range of several orders of magnitude. This general phenomenon was demonstrated in the LPM [49], but is also observed in our calculations in the CPM. This is simply due to the difference in properties of the measure chosen to define the threshold. Note that the ionization yield is determined by factors $\exp\{-\Gamma_i(F)T\}$, (in Eq. 3.2), where widths $\Gamma_i$ are strongly nonlinear functions of $F$. A small increase of the microwave amplitude may strongly increase the resonance width, therefore a necessary decrease of the interaction time, so as to keep their product constant, may be large. Again, it yields another reason why the $T10\%$ seems to be a more sensitive and informative measure of the ionization than the amplitude threshold.

For the sake of further tests of the localization theory predictions, we present the data for stronger fields and higher frequencies in Fig. 5. Again a similar behaviour, i.e., the reasonable classical–quantum predictions agreement at lower frequencies with strong disagreement for higher frequencies, is found. Similar results are observed for the higher microwave amplitude $F_0 = 0.1$ (not shown).

Nevertheless, a sensitive verification of the photonic localization theory requires a comparison of the *quantitative* predictions of the theory with our numerical results. This is difficult, because the photonic localization theory may be used for estimation of the threshold only if the effective threshold is not too high and no significant real ionization occurs (see remarks in the Appendix of [41]). In addition, it is supposed to be valid only for extremely long time, hence $T10\%$ is not a relevant quantity for that purpose.

Fig. 5 brings results for an initial elongated state, $n_0 = 64,\ m_0 = 0$. For an appropriate comparison with the classical simulation, we adjusted $n_{max}$ in such a way that the quantum results for both $n_0 = 48$ and $n_0 = 64$ correspond to the same classical cutoff limit. This is fulfilled when $N = 120$ (160) for $n_0 = 48$ (64), corresponding respectively to $n_{max} = 300$ (resp. 400). Note that typically points of $T10\%$ for $n_0 = 64$ are closer to the classical results than those obtained for $n_0 = 48$. This is expected on the semiclassical grounds and is in agreement with the photonic localization theory which predicts the decrease of the quantum threshold proportional to $n_0^{-1/2}$ [41]. In addition, Fig. 5 shows also data obtained for $n_0 = 48$ and larger $n_{max} = 480$, corresponding to effective ionization cutoff $N \approx 151$. While overall results weakly depend on the cutoff value, for some frequencies discrepancies arise. They typically occur when the shift of the cutoff leads to a change in the number of photons needed for the ionization.

It is a crucial test of the photonic localization theory to check whether the predictions for the localization, Eq. (5.1), do work in the real system. For this purpose, we will focus now on the $F10\%$ ionization threshold which can be determined on the ground of the photonic localization theory and reads [6,41]

$$F10\% = \frac{\omega_0^{7/6}}{\sqrt{8n_0}\mathcal{H}}\sqrt{1-\frac{n_0^2}{N^2}}. \tag{5.4}$$

Recall that $N$ is the principal quantum number corresponding to the effective ionization threshold. For elongated states studied in this subsection, $\tilde{m} = m_0/n_0$ is equal to 0 and $\mathcal{H}$ is just equal to 1. Hence, we are left with 3 parameters: $n_0$, $F_0$ and $\omega_0$. First, we fix $n_0 = 48$ and study the $F10\%$ threshold as a function of $\omega_0$. The result in the high frequency domain is shown in Fig. 6. Apart from the existence of strong fluctuations, there is a general increasing trend with $\omega_0$. The prediction of the photonic localization theory, shown as a solid line, has the same trend but obviously largely overestimates the $F10\%$ threshold. The dotted line, equal to half the theoretical prediction, is in a good overall agreement with the numerical results. Similar results are obtained for a different initial state $n_0 = 64$, as it is also shown in Fig. 6.

In the second test, $\omega_0$ is fixed and the $F10\%$ threshold versus $n_0$ is plotted. For this purpose, we take $\omega_0 = 2.1$ and change the size of the basis such as the ratio $n_0/N$ is kept fixed. The result is shown in Fig. 7. Again, the general trend is well predicted by the photonic localization theory (solid line), but the exact value is significantly different. The dotted line is half of the theoretical prediction, and as before it is in a good agreement with our data. Quite a similar behaviour, with the same factor 0.5, is recorded at frequency $\omega_0 = 2.8$, as it is also shown in Fig. 7.

In the third and final test, we fix both $\omega_0$ and $n_0$ and change only the effective cutoff, $N$, [compare Eq.(5.4)]. The results obtained for $F10\%$ are summarized in Fig. 8. As in the previous figures the prediction of the localization theory significantly overestimates the threshold for large $N$ values, the factor 0.5 improves the agreement considerably. Close inspection of Fig. 6 and Fig. 7 reveals however that the values of $\omega_0$ and $n_0$ chosen to coincide with other figures are not the best. At $\omega_0 = 2.1$ and $n_0 = 48$, $F10\%$ curves in both figures have local maxima, so the factor 0.5 may not work very well. Therefore, we have also attempted to fit the proportionality factor in the $N \to \infty$ limit. The result, represented as a dot-dashed line, agrees with the numerical data quite well.

Note that while both Fig. 6 and Fig. 7 show quite large fluctuations about the average behaviour, these fluctuations are practically absent in Fig. 8. This shows that the origin of fluctuations is not due to the size of the effective cutoff. What is more, the fluctuations are present in the $F10\%$ dependence on $n_0$, (i.e. on the effective $\hbar$), while all classical parameters are kept constant. Thus the fluctuations as a function of $n_0$ are of a quantum origin. We tend to believe that their origin is due to quantum interference between different ionizing paths leading to the continuum and the effect is, to some extent, statistical. Semiclassically the change of $\hbar$ is equivalent to a different level of coarse–graining of the phase space, thus these fluctuations may also reflect transport properties from the initial state to the continuum on a changing, with $n_0$, resolution level. Certainly, observed fluctuations deserve a more detailed study which is, however, beyond the scope of the present paper. Let us only mention here, that similar fluctuations are observed [56] for the individual widths of the well localized, so called wavepacket states [57,58] both for the linear and the circular polarizations.

In conclusion, the photonic localization theory is in a good qualitative agreement with our numerical results for elongated states. Further work is needed to understand the discrepancy of a factor 2. The numerical value of this factor will be dependent on the pulse duration (note that we use pulses of 500 microwave periods) and it will increase for longer interaction times. Since photonic localization theory is valid, strictly speaking, for infinite interaction times, the appropriate correction factor may be larger. Also, in Eq. (5.3), the factor $\mathcal{H}$ is estimated using the $m_0$ value of the initial state. However, during the excitation process, the atom absorbs a few circularly polarized photons and the value of $m_0$ increases. Using a corrected value for $m_0$ in Eq. (5.3) decreases the prediction of the photonic localization theory by approximately 20% and consequently reduces the observed discrepancy.

### B. Angular momentum dependence of the threshold

In most of the experiments [10–12], while the initial principal quantum number, $n_0$, is well defined, the atoms are not preselected with respect to the initial angular momentum quantum numbers. It implies that the threshold is dominated by the states ionizing in the smallest microwave field. In the case of the LPM ionization, these are the elongated states. Do these states ionize first also in the CPM ionization? Classical studies [29] have indicated that elliptical orbits with $\tilde{m} = m_0/n_0 \approx 0.5$ ionize first in a broad range of frequencies.

The quantum test of this prediction is presented in Fig. 9, which shows for a fixed value of $F_0$ and $\omega_0$, the thresholds obtained for all $m_0$ states in the $n_0 = 48$ manifold. Notice a broad minimum around $m_0 \approx n_0/2$ indicating that indeed elliptical states with moderate $m_0$ values are most important for the threshold determination. While the state to state inspection of the data reveals a sensitivity of the ionization efficiency on the choice of $F_0$ and $\omega_0$ values used for the comparison (this is due to the presence of structures in the threshold as a function of $\omega_0$ and the abundance of local maxima in the ionization yield as a function of $F_0$ — compare with Fig. 2), a common overall behaviour is observed in a broad range of both $F_0$ and $\omega_0$.

The increase of the ionization efficiency with $m_0$ for a fixed microwave amplitude and frequency is also predicted by the photonic localization theory [6]. For $\omega_0 \approx 2$, the angular momentum condition for the applicability of the theory is satisfied for almost all $\tilde{m}$ values, up to $\tilde{m}$ close to one (corresponding to circular states). The localization length, Eq. (5.2), is determined by the product of $\mathcal{H}F_0$, where $\mathcal{H}$ monotonically increases with $\tilde{m}$ [compare with Eq. (5.3)]. Thus, for fixed $F_0$ and $\omega_0$, the localization length increases with $\tilde{m}$. A quantitative test is presented in Fig. 10. The solid line is the prediction of the photonic localization theory being also seriously off the numerical data. However, the qualitative trend is well verified (for negative and positive $\tilde{m}$). Correction by a factor 0.5, as before, yields (dashed line) a reasonable agreement with the data. A comparison with Fig. 6 shows that, at $\omega_0 = 2.1$, the fluctuations in the quantum numerical data are not "favorable" and the factor 0.5 does not work too well. Therefore, we have fitted the correction factor so as the numerical value and the modified prediction of the localization theory [6] exactly coincide at $\tilde{m} = 0$. Such a modification yields a dot-dashed line in excellent agreement with numerical data for $|\tilde{m}| < 0.2$. Two important factors are to be noticed. First, the fluctuations are much smaller than in Figs. 6,7. This lack of fluctuations may be explained by correlations between $F10\%$ values for close $m_0$ quantum numbers. The origin of this correlation is mainly classical – the angular momentum in the CPM ionization problem is not conserved and the classical phase space is "shared" by different initial $m_0$ values. The situation is completely different in this respect from a more familiar interaction with the LPM, which conserves $m$. While paths leading to ionization of various $m_0$ states are different, which is reflected by the change of $F10\%$ with $m_0$, some correlations between ionization yields for little differing values of $m_0$ are to be expected. In fact we have verified, by calculating the $n_0$ dependence of $F10\%$ for $m_0 = 2$ (not shown) that the structures observed in Fig. 7 are exactly reproduced for $m_0 = 2$, the only difference being the lower average behaviour, which is consistent with Fig. 10.

Secondly, the localization theory clearly breaks down around $\tilde{m} = 0.5$, whereas the prediction, Eq. (5.1), is supposed to be valid everywhere. Obviously, the conditions in Eq. (5.1) are not sufficiently restrictive. As a matter of fact, the Kepler map relies on the approximation that most of the energy exchange takes place in the vicinity of the perihelion. For states with not–small angular momentum, this cannot be true. Hence, the applicability of the Kepler map is at best restricted to small angular momentum $\tilde{m} \ll 1$ and high frequency.

Our quantum results are, in this respect, in good agreement with a classical study of Nauenberg [20] who, as mentioned in the Introduction, constructed a better, canonical map after noting difficulties with the original Kepler map of Ref. [6]. It would be most interesting, therefore, to compare quantum results with predictions of the photonic localization theory based on the canonical Kepler map of Nauenberg instead of the standard Kepler map of [6]. Note, however, that for a high eccentricity, a small microwave amplitude and a high frequency, the Nauenberg Kepler map is close to the standard Kepler map and no significant difference is likely to take place. On the contrary, at medium and high eccentricities the two maps are significantly different. The breakdown of the photonic localization theory around $\tilde{m} = 0.5$ may be due either to a breakdown of the classical standard Kepler map or a failure of the localization picture itself. This interesting alternative is left for the future studies since a generalization of the theory of [6] for a map of a different functional form is unfortunately not straightforward. The results presented indicate that some effort should be taken in this direction.

### C. Elliptical states

Since states with $m_0 \approx n_0/2$ are "the first to ionize" among different angular momentum states, it is worthwhile to investigate their properties in more detail. As an example, we have chosen the states $n_0 = 48, m_0 = 24$ and $n_0 = 64, m_0 = 32$. Recall that, according to the convention adopted here, positive $\omega_0$ correspond to positive $m$ states in right–polarized microwaves (or $-m$ states in left–polarized microwaves). Fig. 11 presents the moderate frequency behaviour for $n_0 = 48, m_0 = 24$ state. As for the elongated states, one observes the decrease of the threshold with scaled frequency for low frequencies ($\omega_0 \approx 0.6$). For higher frequencies, $T10\%$ is again an increasing (on average) function of $\omega_0$. Note that the microwave amplitudes $F_0$ leading to comparable $T10\%$ for a given scaled frequency are smaller for elliptical states than those used in Fig. 4. This confirms the fact that elliptical states ionize much faster than elongated states in a broad frequency range. A comparison with the corresponding classical simulation reveals again that for effectively weaker pulses, corresponding to longer $T10\%$, the quantum ionization proceeds much slower than its classical counterpart. On the other hand, one may notice, again by comparison between Fig. 4 and Fig. 11, that the classical–quantum disagreement is less pronounced and less sharp for elliptical states than for the elongated states. This is confirmed further in the higher frequency interval – compare Fig. 5 and Fig. 12.

A still higher frequency range is represented in Fig. 13. Note that the overall behaviour of the threshold is quite similar to that observed for lower frequencies (and lower field) in Fig. 12. Namely, there is an apparent agreement between classical and quantum predictions for frequencies at which a given amplitude $F_0$ yields the ionization quite fast, on time scales of the order of few tens of the microwave period. For higher frequencies, the discrepancy between classical and quantum predictions appears. This behaviour, as before, is qualitatively, but not quantitatively (using a similar analysis as for the elongated states) consistent with the photonic localization theory [6].

Finally, Fig. 14 completes the study of elliptical states showing the ionization under the opposite polarization. The comparison of the amplitudes, $F_0$, needed to produce similar $T10\%$ for right–hand and left–hand polarization in a similar absolute frequency range shows the asymmetry between the two possible polarizations. This asymmetry is of a classical origin — a phenomenon like this has been observed in the diffusive regime in the classical studies [29]. Note the strong increase of $T10\%$ for lower frequencies — we are approaching the limit of low frequencies with a relatively high threshold for the left–hand CPM [23,25,31].

It follows naturally that, as discussed on the classical ground in [29], collisions with the nucleus (which are the main classical mechanism for the ionization in the LPM) play little rôle for the CPM in the case of elliptical states with a non–negligible $\tilde{m} = m_0/n_0$ angular momentum. The right polarized microwave acting on the positive $m_0$ states only supplies the electron with additional angular momentum (due to the $\Delta m = +1$ selection rule). Therefore, the electron feels a centrifugal barrier and never comes close to the nucleus. Note, that for the opposite polarization of microwaves, when the angular momentum may diminish, the ionization is much less effective. It confirms the fact that collisions with nucleus play a marginal rôle in the circularly polarized microwave ionization. No singularity of any type is observed in Fig. 9 near $m_0 = 0$, which again proves that collisions with the nucleus are not relevant for the ionization in the CPM.

### D. Circular states

Lastly, let us present results for circular states. As we have seen on the example presented in Fig. 9, these states may ionize at much higher microwave amplitudes than elliptical or elongated states. Thus the results presented below are of no relevance for the ionization threshold of atoms not preselected with respect to the angular momentum. However, efficient methods have been proposed [33] and confirmed experimentally [34] for preparing atoms in initial circular states. One may, therefore, envision a CPM ionization experiment, in which atoms in circular states are prepared before entering the microwave cavity.

It is interesting also to study the CPM ionization from initial circular states for other reasons. While classical studies [29] have shown the diffusive character of the excitation and ionization of such states, Poincaré surface of section studies [20] revealed the presence of regular structures in the phase space even for strong fields. The different character of classical dynamics is visible also in the Chirikov overlap analysis [22], which breaks down for circular orbits. It is interesting to see whether the strongly mixed character of the phase space manifests itself in the behaviour of the quantal ionization thresholds.

Fig. 15 shows $T10\%$ thresholds obtained for different values of $F_0$ for circular states $n_0 = 48, m_0 = 48$ for moderate positive scaled frequencies. Note that, with increasing $F_0$, the curves shift to higher frequencies indicating that the ionization yield decreases (in this frequency range) with $\omega_0$. This agrees with the classical amplitude threshold $F10\%$ behaviour which, for circular states, increases also with the frequency [29]. The dashed lines indicate classical time thresholds and show that the agreement between the classical and quantum predictions is strongly dependent on the time needed for reaching the threshold. For a strong field (at a given frequency), when 10% ionization is reached within up to approximately twenty optical cycles, the classical and quantum thresholds agree quite well. However, the quantum threshold rises much more sharply with the frequency and when the classical diffusion is slow (classical threshold of the order of a thousand cycles), the quantum threshold may exceed the classical value by several orders of magnitude.

The effect looks similar for different frequencies, but occurs simply at different microwave amplitudes. There is no quantitative difference between frequencies less and bigger than $\omega_0 = 1$ — "a starting point" for quantum localization in the LPM studies [6,10]. One should, however, keep in mind that we are considering a situation which is outside the region of applicability of the photonic localization theory [see Eq. (5.1) above]. The observed behaviour may be therefore quite different.

Looking along the vertical axis on the plot, it is clear that, for a given frequency, the changes of $F_0$ will determine whether the agreement between classical and quantum results occurs. For relatively small fields, there will be differences whereas for strong fields — leading to fast diffusion — we will find agreement between classical and quantum results. The region of slow classical diffusion is infested with remnants of regularities observed at still smaller microwave amplitudes (see above). The classical transport in this case goes through several bottlenecks as studied in [59]. Thus a different mechanism leading also to quantum slow–down of the ionization process suggests itself. It has been proposed [60,61] that the finiteness of $\hbar$ leading to the "coarse-graining" of the phase space will "fill" the gaps in broken tori and effectively slow down the quantum excitation leading to the discrepancies between classical and quantum mechanical results (or experimental results) observed in the LPM experiments [60]. Due to the structure of the classical phase space, this explanation of the origin of classical–quantum differences seems to be quite appropriate for the case of the CPM ionization of circular states. However, testing this idea in a numerical experiment is far from being obvious.

A similar behaviour occurs at larger scaled frequencies as shown in Fig. 16 for $F_0 = 0.07$ and in Fig. 17 for $F_0 = 0.1$. Both these figures bring also results for a larger initial principal quantum number $n_0 = 64$. The comparison with corresponding plots for elongated and, in particular elliptical states, reveals that the classical–quantum differences are more pronounced for circular states. This phenomenon may be understood qualitatively in terms of the dimensional argument. Extreme angular momentum states, $|n_0, m_0 = n_0 >$, in right–hand polarized microwaves couple dominantly to $n', m'$ states with large $m'$ (remember the $\Delta m = +1$ selection rule for absorption). Hence the problem becomes effectively one–dimensional. The situation is quite different if $m_0 \ll n_0$ since many more atomic states become coupled together, and we have a truly two–dimensional picture. This may be an argument towards the quantum localization origin of this difference — recall that the Anderson localization is strongly dependent on the system dimensionality. On the other hand, higher density of states for the elliptical initial state (not of all the atomic states, but of those important ones actively involved in the ionization process) means that the effective $\hbar$ becomes smaller. That means that the semiclassical limit is realized faster for the elliptical initial states and the quantum effect of gluing the holes in the regular phase space structures is smaller for these states.

As discussed above, on the average, $T10\%$ increases quite sharply with $\omega_0$ for a given $F_0$ in a whole interval corresponding to classically diffusive motion (i.e. up to $\omega_0 \approx 3 - 3.5$, beyond which the classical ionization thresholds sharply increases). To compare ionization at different frequencies on the same time scale, the amplitude thresholds (for a given pulse duration) are more relevant. For the reasons stated above, we have not tried to determine some unambiguous amplitude thresholds (by, e.g., averaging over the nonmonotonic structures of the type depicted in Fig. 2) but we have checked that in the whole "chaotic" frequency region, the ionization process is quite similar statistically. For this purpose we used the width function, $W(\varphi_0)$ defined as [54]

$$W(\varphi_0) = \exp\left\{-\sum_i \mathcal{R}(<\tilde{\psi}_i|\varphi_0>^2)\ln \mathcal{R}(<\tilde{\psi}_i|\varphi_0>^2)\right\} \tag{5.5}$$

closely related to the entropy and being a measure of the span of the initial state over the Floquet eigenstates. For the same purpose one may simply count the number of Floquet states (ordered in the descending order with respect to their overlap on the initial state) needed to exhaust the norm of $|\varphi_0>$ up to say 99%. For similar values of the ionization probability and for a given pulse length (500 microwave cycles), the width function has similar values in the whole range of frequencies between $\omega_0 = 0.6$ and 2.6 and is a strongly increasing function of the ionization probability value chosen for comparison. In particular, $W \approx e^3$ for a 20% ionization during 500 microwave cycles corresponding to about 80 Floquet states needed to exhaust the norm of the initial state. For 1% ionization, we obtain $W \approx e^{1.5}$ with roughly 15 important Floquet states (note that $W$ function, while being roughly proportional to the number of important Floquet states in the expansion of the initial state, underestimates this number quite strongly). These numbers start to decrease significantly only for high frequencies when we come close to the classically forbidden region.

Finally, let us discuss the case of oppositely polarized microwaves presented in Fig. 18 as a function of "negative" frequency (see the discussion above). One observes a very fast increase of the threshold time with absolute frequency value, even sharper than that demonstrated in Fig. 15. However, the microwave amplitudes used are roughly *one order of magnitude* higher than the corresponding values needed for right–polarized microwaves in the same frequency interval. This reflects the strong asymmetry between right–hand and left–hand polarized microwaves. Comparing the values of the field necessary to obtain reasonable $T10\%$ (see also Fig. 14), one notices that the asymmetry is much more pronounced for circular than for elliptical states (as expected, since there should be no asymmetry of this kind for elongated states). Let us repeat that the effect (occurring at high frequencies, in contrast to the asymmetry of another origin at low frequencies [23,25,31]) is of a purely classical origin and has been discussed in detail in [29]. Also note a very good agreement between the classical and quantum predictions for $\omega_0 < 1$.

## VI. CLASSICALLY SCALING STRUCTURES

Note that, while the classical time thresholds are quite smooth, the quantum results reveal several frequency-dependent structures. Some of them may be associated with incidental intermediate quantum resonances or the threshold effects (with increasing frequency less and less photons are required for ionization). These structures would not however scale classically. Therefore, they would occur at different values of the scaled frequency $\omega_0$.

The comparison of the thresholds obtained for different $n_0$ reveals, on the other hand, a number of classically scaling structures, i.e., resonant features centered on the *scaled* frequency $\omega_0$. Arguably the most prominent example appears for the circular states, note the structure present at frequency $\omega_0 = 2.8$ in Fig. 17. This local inhibition of ionization (manifested by the corresponding increase of $T10\%$) persists also for different "effective ionization thresholds" — compare results for $n_0 = 48$ and $N = 151$. Interestingly, even classical simulations show a slight slow down of the ionization time in the vicinity of that frequency confirming the classical nature of the resonance (although the effect is much weaker classically). This is in contrast to LPM studies where quantal results revealed classically scaling structures not visible in classical simulations. Apparently the classical phase space structures in the CPM ionization of circular states are stronger (i.e. the motion more regular) than for the LPM case. For lower microwave amplitudes (compare Fig. 16) in the vicinity of $\omega_0 = 2.8$, the ionization is practically totally suppressed for both $n_0 = 48$ and $n_0 = 64$.

Fig. 16 yields another example. Note the locally increased $T10\%$ at $\omega_0 = 2.1$ appearing for $F_0 = 0.025$ for both $n_0 = 48$ and $n_0 = 64$ circular states. The presence of such structures is dependent on $F_0$ — it disappears for higher amplitude $F_0 = 0.04$ when $T10\%$ is much smaller. Such a behaviour is consistent with the proposed classical origin of the structure — for stronger microwave amplitude the remnants of regularity in the classical phase space become weaker and less significant.

Similarly, classically scaling structures appear for elliptical states, as exemplified in Fig. 12 and Fig. 13 or for elongated states (compare Fig. 5). In the latter case, however, a slight shift in $\omega_0$ of the resonance position with $n_0$ may be observed. As before, we believe that the observed structures are of the classical origin and are related to structures in the phase space, or more precisely, they correspond to localization of the Floquet states (contributing to the expansion in Eq. (3.2) in the dominant way) on the classical phase space structures. The detailed study of the localization properties of Floquet states is beyond the scope of this paper and will be studied in [62].

## VII. SUMMARY AND CONCLUSIONS

We have presented extensive results for the ionization threshold dependence of the two–dimensional hydrogen atom illuminated by circularly polarized microwaves at realistic initial principal quantum numbers $n_0 \approx 50-60$ and for different initial angular momenta, $m_0$. The results have been compared, whenever it was possible [i.e., for parameters within limits given by Eq. 5.1], with the predictions of the photonic localization theory [6,41]. For elongated, high eccentricity states, quite nice qualitative agreement of our numerical results with that theory has been observed. However, theoretical predictions for the $F10\%$ threshold are systematically 2 times larger that the numerical results. In our opinion, a revision of the localization theory is needed. The comparison of the ionization threshold dependence on the angular momentum with the predictions of the theory based on the Kepler map has shown significant differences which, in our opinion, suggest that the regime of validity of the Kepler map approach and the photonic localization theory is more limited in the parameter space then anticipated originally [6]. This is in agreement with the classical discussion presented by Nauenberg [20]. His improved, canonical Kepler map does not allow to be incorporated easily into the photonic localization theory in order to obtain the corrected quantum predictions. Such a development of the theory is, therefore, badly needed.

We have shown also, by a detailed study of classical–quantum correspondence for cir-cular states (for which the photonic localization theory cannot be applied), the presence of a strong, localization-like behaviour in that case. Due to the mixed character of the phase space for initial circular states [20,22] the origin of this localization is argued to be qualitatively explained by a quantum enhanced slow–down of the diffusion by remnants of regularities — the picture originally developed for the LPM ionization [60,61]. This is in agreement with the fact that the presence of classical — quantum differences is strongly dependent on the time scale on which the ionization occurs and only weakly on the frequency of the CPM. Note that such a picture can coexist with the photonic localization point of view. Further work is needed to clarify these points.

The asymmetry between thresholds observed for the left–hand polarized and right–hand polarized radiation confirmed that the collisions with nucleus play minor rôle in the CPM ionization of elliptical and circular states.

The explanation of classical–quantum discrepancies as resulting from a quantum suppression of classical diffusion due to a partial localization on remnants of classical regularities is further confirmed by the observed several classically scaling structures. We leave out to a subsequent study a detailed analysis of these structures as well as the discussion of localization properties of important Floquet states in such cases. Similarly, in a separate study, we will provide a detailed analysis of wavepackets stable against ionization, observed close to the 1 : 1 stable resonant island [57], a preliminary study of which has been already presented elsewhere, [58].

We have profited a lot from discussions with Andreas Buchleitner and Dima Shepelyansky. Laboratoire Kastler Brossel, de l'Ecole Normale Superieure et de l'Université Pierre et Marie Curie, is unité associée 18 du CNRS. CPU time on a Cray C98 and on a Cray YMP-EL has been provided by IDRIS and by the Centre de Calcul pour la Recherche de l'Université Pierre et Marie Curie. This work was supported by the Polish Committee of Scientific Research under grant No. 2P302 102 06 (J.Z. and R.G.).

## REFERENCES

[1] see, e.g., D. Delande in *Les Houches Session LII, Chaos and Quantum Physics 1989*, eds. M.-J. Giannoni, A. Voros and J. Zinn-Justin, (North-Holland, Amsterdam) 1991, p.665.

[2] J. Zakrzewski, K. Dupret, and D. Delande, Phys. Rev. Lett. **74**, 522 (1995).

[3] J. E. Bayfield and P. M. Koch, Phys. Rev. Lett. **33**, 258 (1974).

[4] J. G. Leopold and I. C. Percival, J. Phys. B**12**, 709 (1979).

[5] G. Casati, B. V. Chirikov, D. L. Shepelyansky, and I. Guarneri, Phys. Rep. **154**, 77 (1987).

[6] G. Casati, I. Guarneri, and D. L. Shepelyansky, IEEE J. Quant. Electron. **24**, 1420 (1988).

[7] B. V. Chirikov in *Les Houches Session LII, Chaos and Quantum Physics 1989*, eds. M.-J. Giannoni, A. Voros and J. Zinn-Justin, (North-Holland, Amsterdam) 1991, p.443.

[8] R. V. Jensen, S. M. Susskind, and M. M. Sanders, Phys. Rep. **201**, 1 (1991).

[9] A. Buchleitner, Thèse de Doctorat, Université Pierre et Marie Curie (in French), unpublished, Paris (1993).

[10] P. M. Koch, in Proc. of Eighth South African Summer School in Theoretical Physics: Chaos and Quantum Chaos, 13-24 January 1992, Blydepoort, Eastern Transvaal, Republic of South Africa (Springer-Verlag 1993).

[11] P. M. Koch and K. A. H. van Leeuwen, Phys. Rep. **255**, 289 (1995).

[12] J. E. Bayfield and D. W. Sokol, in Physics of Atoms and Molecules: Atomic Spectra and Collisions in External Fields, ed. K.T. Taylor, M. H. Nayfeh, and C. W. Clark, (Plenum, New York, 1988).

[13] M. V. Berry and K. E. Mount, Rep. Prog. Phys. **35**, 315 (1972).

[14] J. Mostowski and J. J. Sanchez-Mondragon, Opt. Commun. **29**, 293 (1979).

[15] B. I. Meerson, E. A. Oks, and P. V. Sasorov, J. Phys. B **15**, 3599 (1982).

[16] P. Fu, T. J. Scholz, J. M. Hettema, and T. F. Gallagher, Phys. Rev. Lett. **64**, 511 (1990).

[17] T. F. Gallagher, Mod. Phys. Lett. B **5**, 259 (1991).

[18] P. M. Koch, private communication.

[19] M. Nauenberg, Phys. Rev. Lett. **64**, 2731 (1990).

[20] M. Nauenberg, Europhys. Lett. **13**, 611 (1990).

[21] J. A. Griffiths and D. Farrelly, Phys. Rev. A **45**, R2678 (1992).

[22] J. E. Howard, Phys. Rev. A **46**, 364 (1992); *ibid.* **51**, 3934 (1995).

[23] K. Rzążewski and B. Piraux, Phys. Rev. A**47**, R1612 (1993).

[24] P. Kappertz and M. Nauenberg, Phys. Rev. A**47**, 4749 (1993).

[25] D. Delande, R. Gębarowski, M. Kuklińska, B. Piraux, K. Rzążewski, and J. Zakrzewski, Super–Intense Laser–Atom Physics, (Proc. of NATO Advanced Workshop SILAP III, Han-sur-Lesse, Belgium, January 1993), ed. B. Piraux, A. L'Huillier, and K. Rzążewski (Plenum, New York), p.317.

[26] R. Gębarowski and J. Zakrzewski, Phys. Rev. A**50** 4408(1994).

[27] M.J. Raković and S.-I. Chu, Phys. Rev. A**50**, 5077 (1994); *ibid.* **52**, 1358 (1995).

[28] D. Farelly and T. Uzer, Phys. Rev. Lett. **74**, 1720 (1995).

[29] R. Gębarowski and J. Zakrzewski, Phys. Rev. A**51**, 1508 (1995).

[30] The paper [29] contains the untrue claim of using for the first time a full regularization

of Coulomb singularity in the microwave ionization problem. We would like to use this opportunity to point out that the same approach was used earlier by Griffiths and Farrely for the same problem [21] and introduced in the context of magnetized hydrogen atom by O. Rath and D. Richards, J. Phys. B**21**, 555 (1988).

[31] J. Zakrzewski, D. Delande, J. C. Gay, and K. Rzążewski, Phys. Rev. A**47**, R2468 (1993).

[32] D. Wintgen, Z. Phys. D **18**, 125 (1991).

[33] D. Delande and J. C. Gay, Europhys. Lett. **5**, 303 (1988).

[34] J. Hare, M. Gross, and P. Goy, Phys. Rev. Lett. **61**, 1938 (1988);

[35] G. Floquet, Ann. Ecole Norm. Sup. **12**, 47 (1883); J. H. Shirley, Phys. Rev. **138**, B979 (1965).

[36] F. V. Bunkin and A. Prokhorov, Sov. Phys. JETP **19**, 739 (1964).

[37] A. Franz, H. Klar, J. T. Broad, and J. S. Briggs, J. Opt. Soc. Am. B **7**, 545 (1990).

[38] T. P. Grozdanov, M. J. Raković, and E. A. Solov'ev, J. Phys. **B25**, 4455 (1992).

[39] D. Delande, Thèse de doctorat d'état, Université de Paris, 1988 (in French) *unpublished.*

[40] R. Gębarowski, PhD Thesis (in Polish), Jagiellonian University, Kraków, Poland (1995) *unpublished.*

[41] G. Casati, I. Guarneri, and D. L. Shepelyansky, Physica A**163**, 205 (1990).

[42] F. Benvenuto, G. Casati, and D. L. Shepelyansky, Phys. Rev. A**45**, R7670 (1992); A**47**, R786 (1993).

[43] Obviously $n_A + n_B$ must be even. The eigenbasis of 2D hydrogen in $u, v$ coordinates is a subset of the full Hilbert space of the isotropic 2D harmonic oscillator. This is related to the properties of the dynamical symmetries of the isotropic 2D oscillator discussed in detail, e.g., in [39].

[44] Note that once the widths of the state followed in the adiabatic-diabatic approach described above have been found for a range of $F$ values between $F = 0$ and some $F = F_{max}$ ionization field dependence may be studied, using Eq. (3.1) for and arbitrary pulse shape of the maximal amplitude up to $F_{max}$.

[45] W. P. Reinhardt, Ann. Rev. Phys. Chem. **33**, 323 (1982); Y. K. Ho, Phys. Rep. **99**, 1 (1983).

[46] S–I Chu and W. P. Reinhardt, Phys. Rev. Lett. **39**, 1195 (1977).

[47] A. Maquet, S–I. Chu, W.P. Reinhardt, Phys. Rev. A**27**, 2946 (1983).

[48] R.M. Potvliege and R. Shakeshaft, Phys. Rev. A**38**, 4597 (1988). M. Dörr, R.M. Potvliege, D. Proulx, and R. Shakeshaft, Phys. Rev. A**43**, 3729 (1991).

[49] A. Buchleitner and D. Delande, Chaos, Solitons, and Fractals, **5**, 1125 (1995).

[50] A. Buchleitner, D. Delande, and J.–C. Gay, J. Opt. Soc. Am. B **12**, 505 (1995).

[51] M. Arndt, A. Buchleitner, R. N. Mantegna, and H. Walther, Phys. Rev. Lett. **67**, 2435 (1991).

[52] D. Delande, A. Bommier, J. C. Gay, Phys. Rev. Lett. **66**, 141 (1991).

[53] J. Zakrzewski and D. Delande, Phys. Rev. E**47**, 1665 (1993).

[54] R. Blümel and U. Smilansky, Z. Phys. D**6**, 83 (1987).

[55] H. P. Breuer, K. Dietz, and M. Holthaus, Z. Phys. D**18**, 239 (1991).

[56] J. Zakrzewski, D. Delande, and A. Buchleitner, Phys. Rev. Lett. (1995) in press.

[57] I. Bialynicki-Birula, M. Kaliński, and J. H. Eberly, Phys. Rev. Lett. **73**, 1777 (1994).

[58] D. Delande, J. Zakrzewski, and A. Buchleitner, Europhys. Lett. **32**, 107 (1995).

[59] R. S. MacKay, J. D. Meiss, I. C. Percival, Physica **13D**, 55 (1984).

[60] R. S. MacKay and J. D. Meiss, Phys. Rev. A**37**, 4702 (1988).

[61] J. D. Meiss, Phys. Rev. Lett. **62**, 1576 (1989).

[62] J. Zakrzewski and D. Delande, *in preparation.*

## FIGURES

FIG. 1. Scaled 10% ionization threshold field amplitude, $F10\%$ versus scaled frequency $\omega_0$ of the circularly polarized microwave (CPM) for the circular state, $|24, 24>$. Open squares connected by dotted line give the classical prediction for a smooth pulse of duration $T = 500$ microwave cycles with a smooth turn on/off over 25 cycles. Filled circles represent quantum results of the single Floquet state approximation (SFSA) for the same pulse, filled triangles have been obtained using the rectangular pulse approximation (RPA), of duration $T = 500$.

FIG. 2. Probability of ionization at scaled frequency $\omega_0 = 1$ as a function of the scaled microwave amplitude $F_0$ for the circular state $|24, 24>$ for the duration of the rectangular pulse $T = 500$ microwave cycles (filled circles connected by solid line), $T = 2000$ (dashed line) and $T = 5000$ dotted line. Note the abundance of local maxima of the ionization curve as a function of $F_0$.

FIG. 3. Comparison of SFSA (open triangles) and RPA (filled circles) for the circular state $|24, 24>$ in the $T10\%$ (time) threshold versus scaled frequency $\omega_0$ at scaled microwave field $F_0 = 0.1$. Note the strong discrepancy between the two approaches for low frequencies (short times) with better agreement for high frequencies and long $T10\%$.

FIG. 4. Length of the rectangular pulse (in microwave cycles) leading to 10% ionization, i.e., the $T10\%$ threshold (according to the definition adopted in the text) versus the scaled frequency, $\omega_0$ for the initial elongated state $n_0 = 48, m_0 = 0$ illuminated by a circularly polarized microwave, for different scaled amplitudes $F_0$. Filled symbols connected by solid lines (to guide the eye) represent the quantum results for the "effective ionization threshold", $N = 120$. Open symbols connected by dashed lines give results of classical simulations corresponding to the same effective threshold. Triangles, circles, and diamonds correspond, respectively to $F_0 = 0.025, 0.03, 0.04$.

FIG. 5. $T10\%$ thresholds for $F_0 = 0.05$ and elongated state $n_0 = 64, m_0 = 0$ (triangles connected by line), $n_0 = 48, m_0 = 0$ (filled circles connected by line), open circles connected by dot-dashed line correspond to threshold obtained for a higher effective cutoff. The results of classical simulations are represented by open small squares connected by dotted line. Possible classically scaling structures appear at $\omega_0 \approx 2.1$ and $\omega_0 \approx 2.7$.

FIG. 6. $F10\%$ threshold for the elongated state $n_0 = 48, m_0 = 0$ ($n_0 = 64, m_0 = 0$) represented by filled circles (resp. open triangles) as a function of the scaled field frequency (interaction time 500 field cycles). The solid (resp. dashed) line is the prediction of the photonic localization theory [6] and the dotted (resp. dot-dashed) line half this value. The localization theory predicts the correct functional dependence, but with a wrong coefficient.

FIG. 7. $F10\%$ threshold for the elongated state $n_0, m_0 = 0$ as a function of the principal quantum number $n_0$ (interaction time 500 field cycles). Filled circles (resp. triangles) represent the data for scaled frequency $\omega_0 = 2.1$ (resp. $\omega_0 = 2.8$). The predictions of the photonic localization theory [6] are shown by solid (resp. dashed) lines while the dotted (resp. dot-dashed) lines give half of the theoretical prediction. The localization theory predicts the correct functional dependence, note that the "correction factor" is the same for both frequencies.

FIG. 8. The dependence of the $F10\%$ threshold on the effective ionization threshold $N$ for $n_0 = 48$ at $\omega_0 = 2.1$(filled circles). The prediction of the photonic localization theory [41] is shown by solid line, dashed line gives half of this value. Dot–dashed line is a result of a fit of the constant factor in the localization theory prediction to numerical value for large $N$ and in excellent way describes variation of $F10\%$ threshold for the whole interval of $N$ values shown.

FIG. 9. The dependence of the $T10\%$ threshold for ionization on the initial angular momentum $m_0$ for the $n_0 = 48$ manifold. The filled (resp. open) triangles correspond to thresholds obtained at $\omega_0 = 2.1$ and $F = 0.04$ (resp. $F = 0.05$) while circles present the data for $F = 0.04$ and a smaller frequency $\omega_0 = 1.9$. The lowest threshold values are obtained for $m_0 \approx n_0/2$.

FIG. 10. The dependence of the $F10\%$ threshold for ionization on the initial angular momentum $m_0$ for the $n_0 = 48$ manifold at $\omega_0 = 2.1$. The prediction of the photonic localization theory [6] is shown by solid line, dashed line gives half this value. Dot-dashed line is a result of a fit of the constant factor in the localization theory prediction to numerical value at $m_0 = 0$ and in excellent way describes variation of $F10\%$ threshold for $|m_0| < 10$ (corresponding to scaled angular momentum $|\tilde{m}| < 0.2$.

FIG. 11. $T10\%$ threshold for elliptical state $n_0 = 48, m_0 = 24$. The filled symbols correspond to quantum results, while the small open symbols stand for classical simulations. $F_0 = 0.015, 0.025$ and $F_0 = 0.04$ data are represented by squares, circles and triangles, respectively. The large open circles connected by a dot-dashed line correspond to a higher lying initial state $n_0 = 64, m_0 = 32$ – comparison with filled circles allows for locating classically scaling structures.

FIG. 12. Same as Fig. 11 but for $F_0 = 0.05$. Filled circles – $n_0 = 64, m_0 = 32$; open circles – $n_0 = 48, m_0 = 24$, diamonds – $n_0 = 48, m_0 = 24$ for higher effective cutoff, squares connected by a dotted line - classical simulation results.

FIG. 13. Same as Fig. 12 but for $F_0 = 0.1$. Filled circles – $n_0 = 64, m_0 = 32$; open circles – $n_0 = 48, m_0 = 24$, diamonds – $n_0 = 48, m_0 = 24$ for higher effective cutoff, squares connected by a dotted line — classical simulation results.

FIG. 14. $T10\%$ threshold for the elliptical state $n_0 = 48, m_0 = 24$ in left-hand CPM (or $m_0 = -24$ in right-hand polarized radiation). The filled (resp. open) symbols connected by lines give the quantum (resp. classical) results. The circles, diamonds and triangles correspond to $F_0 = 0.04, 0.05$ and $F_0 = 0.1$, respectively.

FIG. 15. $T10\%$ threshold for the initial circular state $n_0 = 48, m_0 = 48$ illuminated by a right-hand CPM for different scaled amplitudes $F_0$. The filled (resp. open) symbols connected by lines give the quantum (resp. classical) results. The squares, circles, diamonds and triangles represent the data obtained for $F_0 = 0.02,\ 0.03,\ 0.04$ and $0.07$, respectively.

FIG. 16. $T10\%$ threshold for $F_0 = 0.07$ obtained for right–hand CPM ionization of circular state $n_0 = 48, m_0 = 48$ (filled triangles, cutoff $N = 120$) and $n_0 = 64, m_0 = 64$ (filled circles, cutoff $N = 160$).

FIG. 17. Same as Fig. 16, but for stronger amplitude, $F_0 = 0.1$. The open diamonds represent the quantum data obtained also for $n_0 = 48, m_0 = 48$ state, but for higher effective cutoff, $N \approx 151$ - for some frequencies the shift of the cutoff strongly affects the value of $T10\%$. Note the classically scaling structure at $\omega_0 \approx 2.8$. Classical prediction is given by open circles connected by dotted line.

FIG. 18. Same as Fig. 16, but for left-polarized CPM (or right-polarized CPM for $n_0 = 48, m_0 = -48$ state). Filled triangles – $F_0 = 0.15$, filled circles – $F_0 = 0.3$, the corresponding classical simulations are represented by open symbols. The required microwave amplitudes to obtain a reasonable T10% values are much larger than those corresponding to right-polarized CPM - compare figures above.

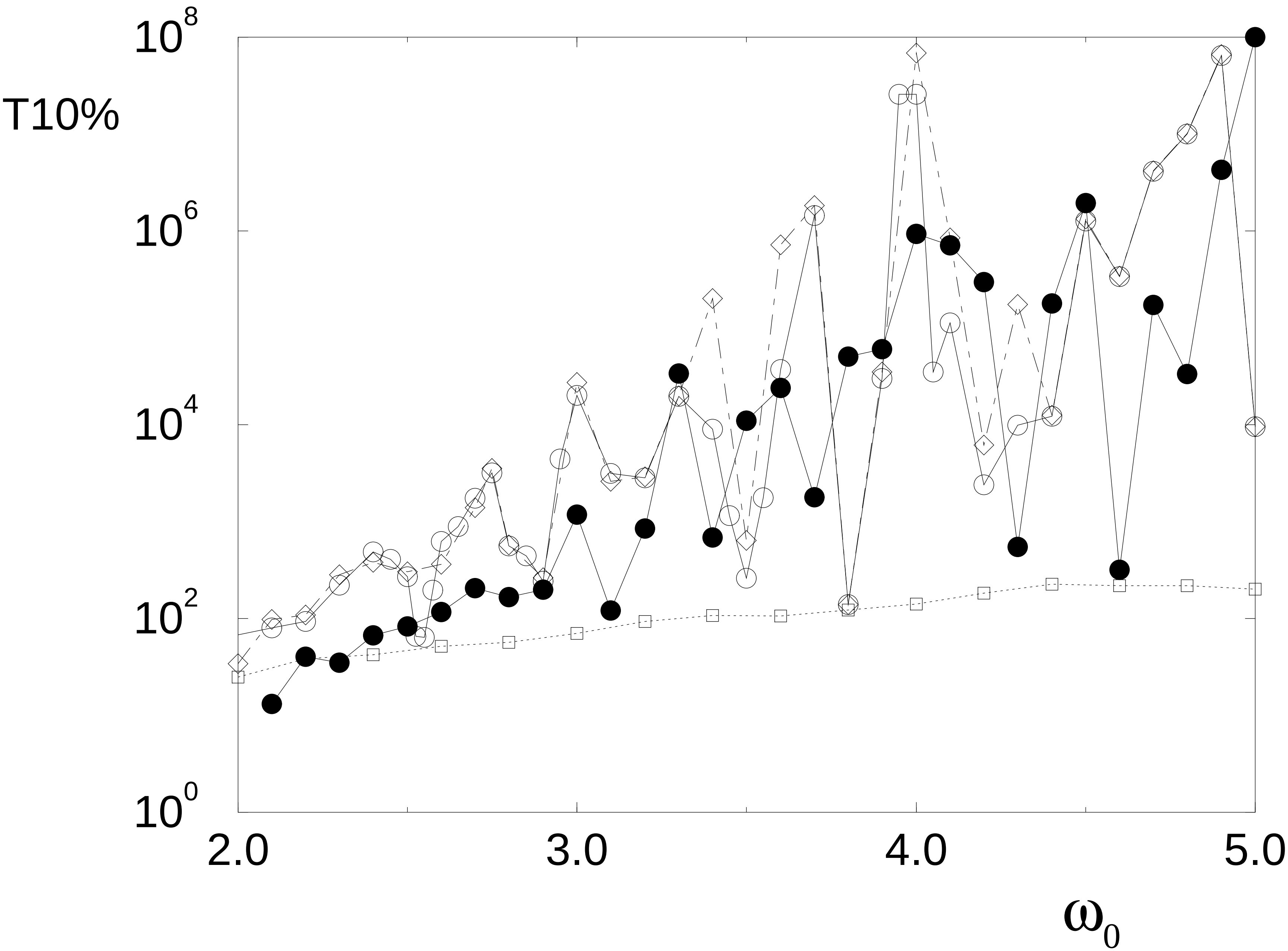

T10%
$10^8$
$10^6$
$10^4$
$10^2$
$10^0$
2.0
3.0
4.0
5.0
$\omega_0$

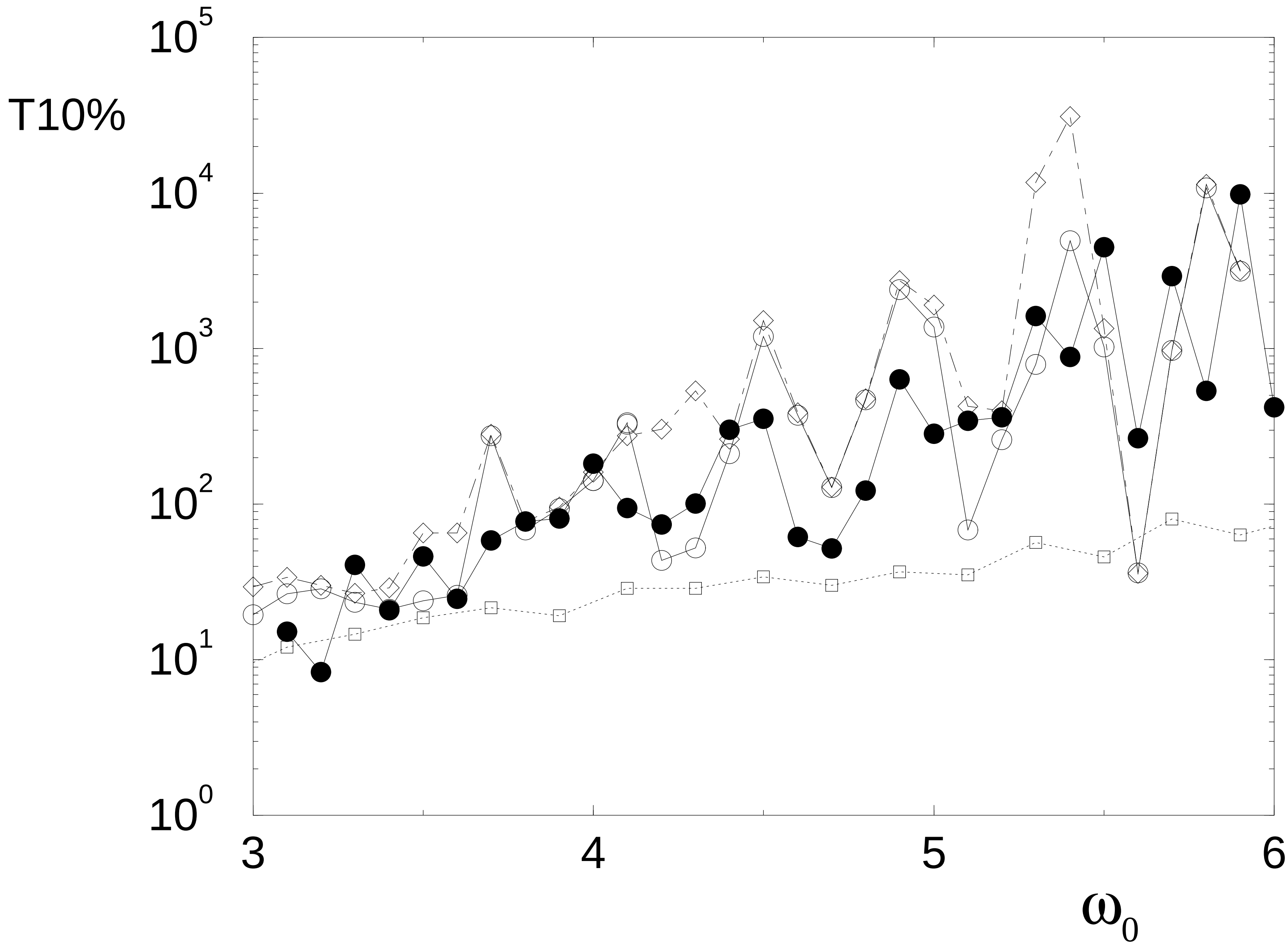

T10%
$10^5$
$10^4$
$10^3$
$10^2$
$10^1$
$10^0$
3
4
5
6
$\omega_0$

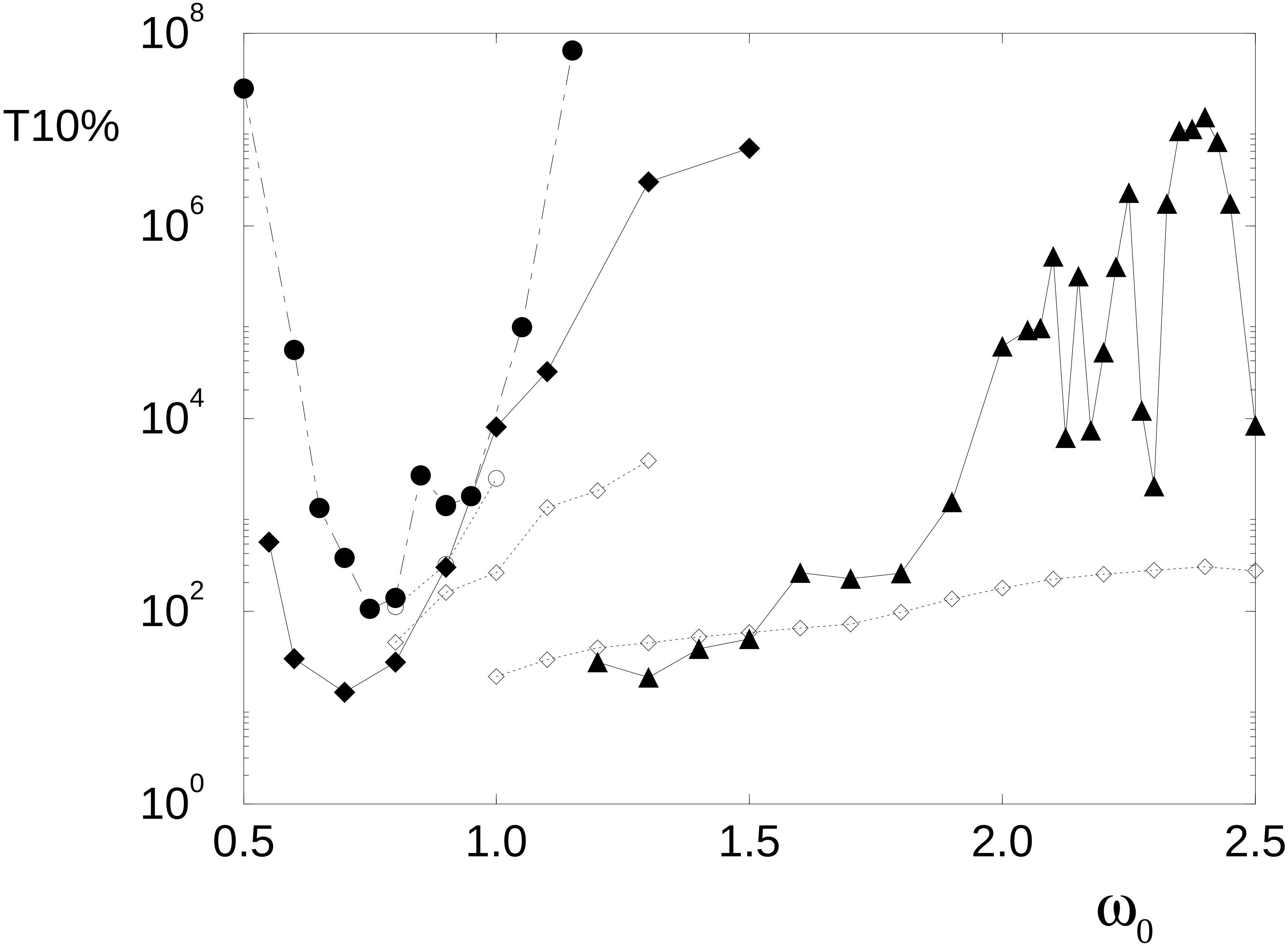

T10%
10^8
10^6
10^4
10^2
10^0
0.5
1.0
1.5
2.0
2.5
ω0

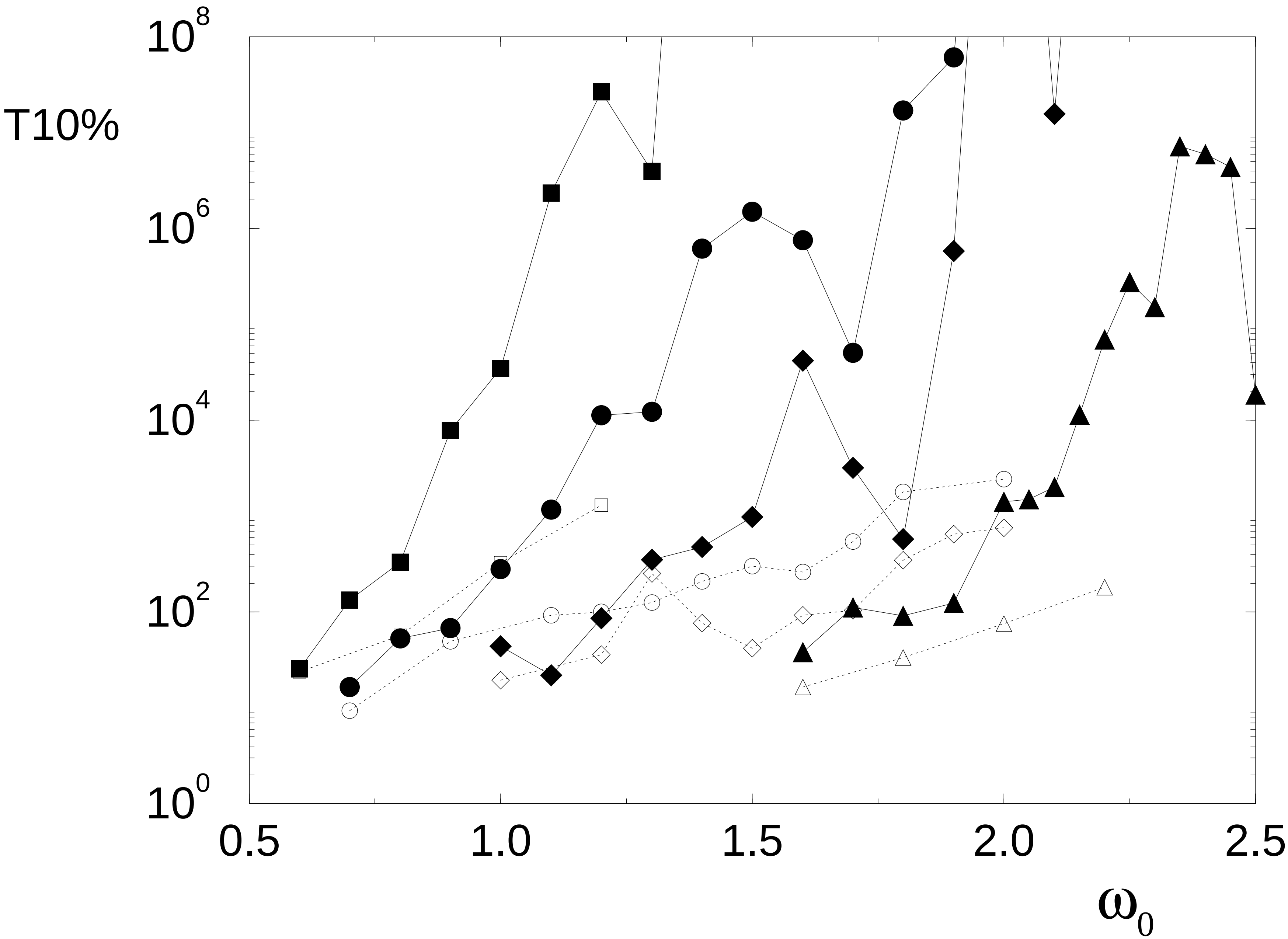

T10%
10^8
10^6
10^4
10^2
10^0
0.5
1.0
1.5
2.0
2.5
ω0

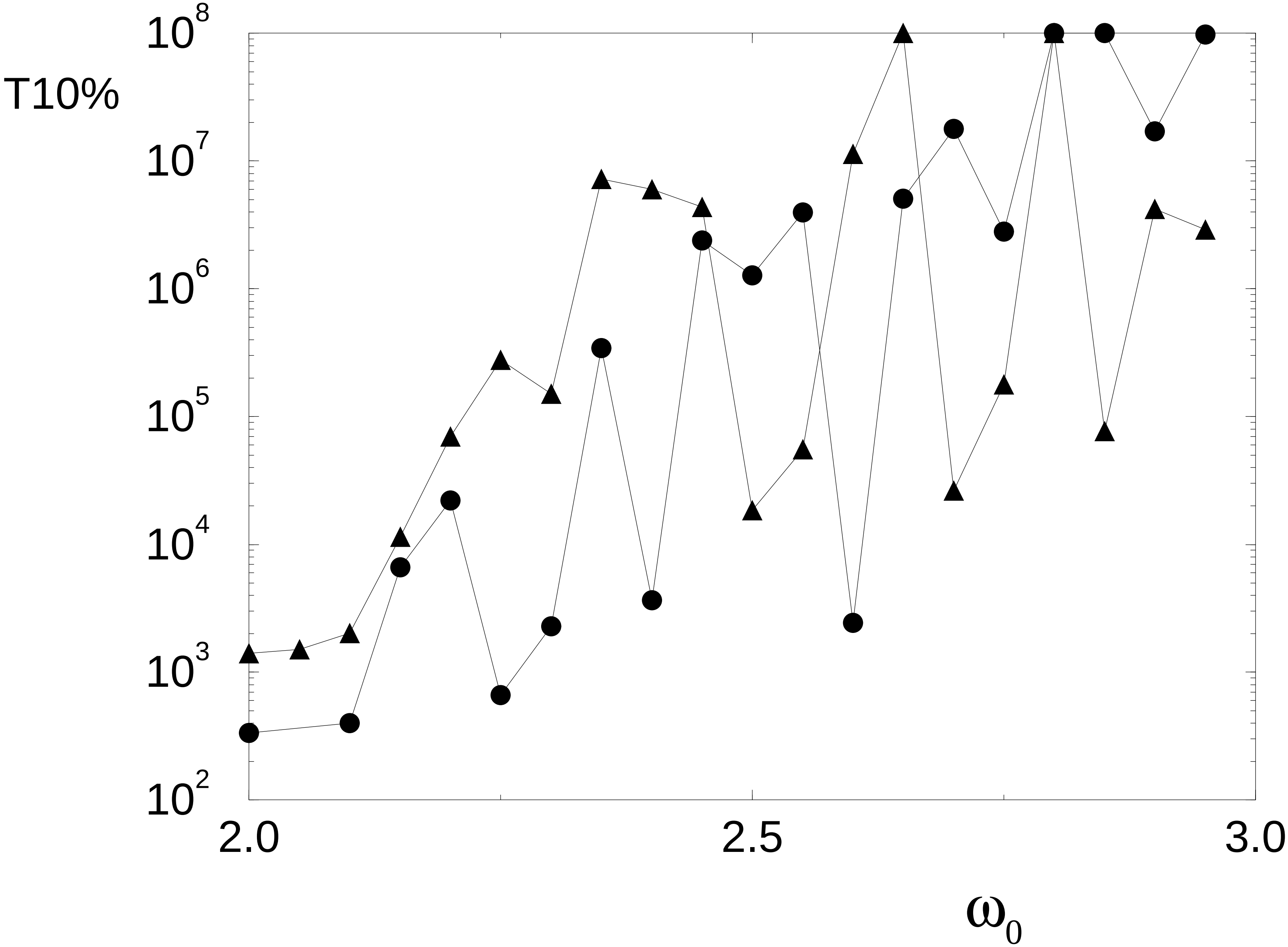
T10%
$10^8$
$10^7$
$10^6$
$10^5$
$10^4$
$10^3$
$10^2$
2.0
2.5
3.0
$\omega_0$

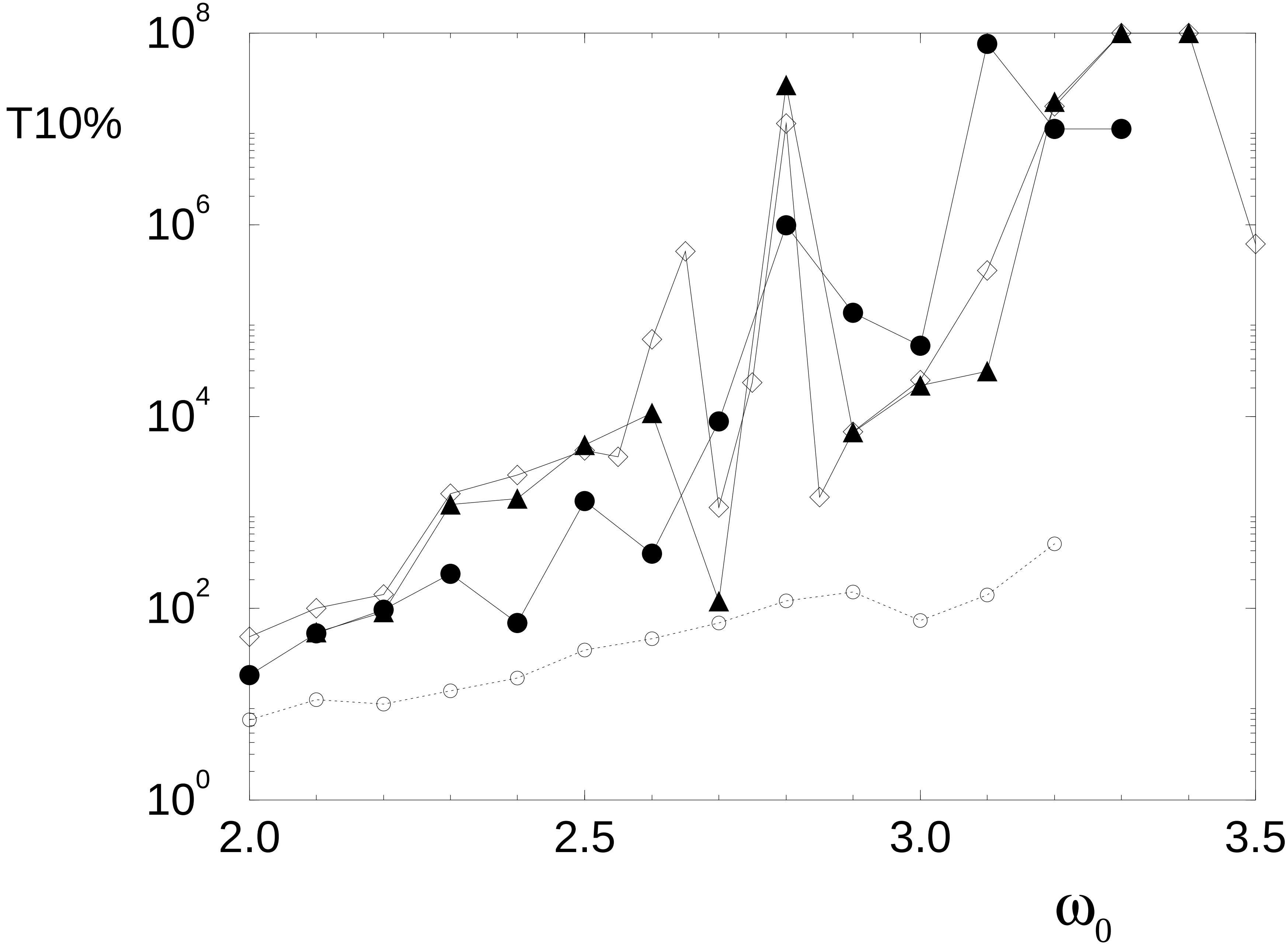

T10%
10^8
10^6
10^4
10^2
10^0
2.0
2.5
3.0
3.5
ω0

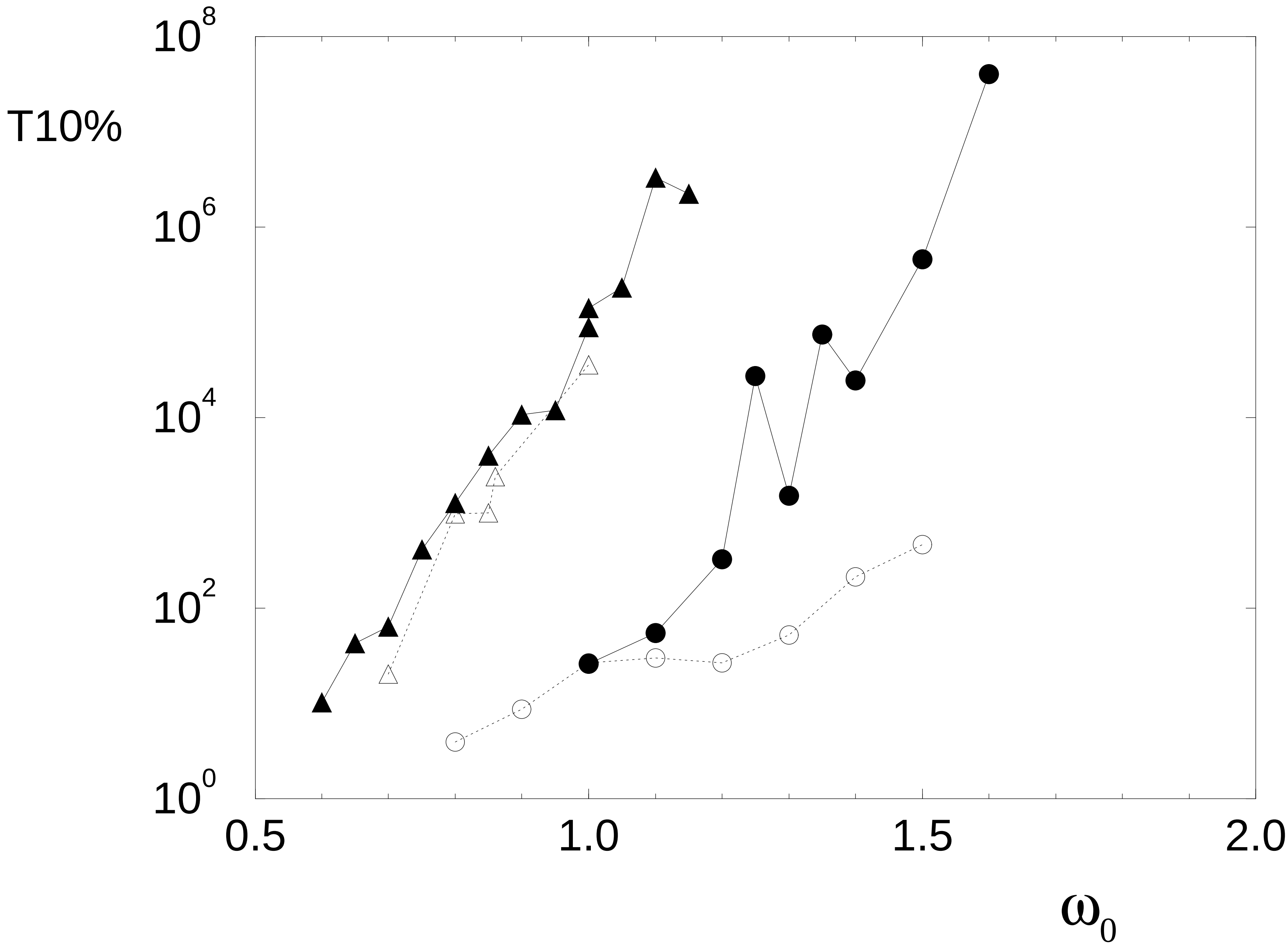
T10%
10^8
10^6
10^4
10^2
10^0
0.5
1.0
1.5
2.0
ω0